\documentclass{emulateapj}

\newcommand{\myemail}{ymiki@ccs.tsukuba.ac.jp}

\slugcomment{}
\shorttitle{Properties of the Andromeda stellar stream and its progenitor}
\shortauthors{Miki, Mori, \& Rich}
\begin{document}


\title{Collision tomography: Physical properties of possible progenitors of the Andromeda stellar stream}
%%%%%%%%%%%%%%%%%%%%%%%%%%%%%%%%%%%%%%%%%%%%%%%%%%%%%%%%%%%%%%%%%%%%%%%%%%%%%%
%%%%%%%%%%%%%%%%%%%%%%%%%%%%%%%%%%%%%%%%%%%%%%%%%%%%%%%%%%%%%%%%%%%%%%%%%%%%%%


\author{Yohei Miki$^{1}$, Masao Mori$^{1}$, and R. Michael Rich$^{2}$}
\affil{$^1$Center for Computational Sciences, University of Tsukuba, 
  Tsukuba, Ibaraki 305-8577, Japan; 
  \myemail}
\affil{$^2$Department of Physics and Astronomy, University of California, 
  Los Angeles, CA 90095-1562, USA}
%%%%%%%%%%%%%%%%%%%%%%%%%%%%%%%%%%%%%%%%%%%%%%%%%%%%%%%%%%%%%%%%%%%%%%%%%%%%%%
%%%%%%%%%%%%%%%%%%%%%%%%%%%%%%%%%%%%%%%%%%%%%%%%%%%%%%%%%%%%%%%%%%%%%%%%%%%%%%

\begin{abstract}
To unveil a progenitor of the Andromeda Giant Stellar Stream, we investigate the interaction between an accreting satellite galaxy and the Andromeda Galaxy using an $N$-body simulation. 
A comprehensive parameter study with 247 models is performed by varying size and mass distribution of the progenitor dwarf galaxy. 
We show that the binding energy of the progenitor is the crucial parameter in reproducing the Andromeda Giant Stellar Stream and the shell-like structures surrounding the Andromeda Galaxy. 
As a result of the simulations, the progenitor must satisfy a simple scaling relation between the core radius, the total mass and the tidal radius. 
Using this relation, we successfully constrain the physical properties of the progenitors to have mass ranging from $5\times10^8 M_\odot$ to $5\times10^9 M_\odot$ and central surface density around $10^3\, M_\odot\, \mathrm{pc}^{-2}$. 
A detailed comparison between our result and the nearby observed galaxies indicates that possible progenitors of the Andromeda Giant Stellar Stream include a dwarf elliptical galaxy, a dwarf irregular galaxy, and a small spiral galaxy.
\end{abstract}
%%%%%%%%%%%%%%%%%%%%%%%%%%%%%%%%%%%%%%%%%%%%%%%%%%%%%%%%%%%%%%%%%%%%%%%%%%%%%%
%%%%%%%%%%%%%%%%%%%%%%%%%%%%%%%%%%%%%%%%%%%%%%%%%%%%%%%%%%%%%%%%%%%%%%%%%%%%%%


\section{Introduction}
\label{section:introduction}
%%%%%%%%%%%%%%%%%%%%%%%%%%%%%%%%%%%%%%%%%%%%%%%%%%%%%%%%%%%%%%%%%%%%%%%%%%%%%%
According to the hierarchical model of galaxy formation, minor mergers and accretion events have played a crucial role in the formation of currently observed large galaxies such as the Milky Way and M31 (commonly known as the Andromeda Galaxy). 
This hypothesis is supported by the discovery of tidal features such as the Sagittarius stream and the giant stellar stream in M31. 
Furthermore, photometric and spectroscopic observations of the spatial distribution and radial velocity distribution of red giant stars and of the metallicity distribution near these galaxies have revealed other substructures \citep{fer02, iba04, iba05, guh06, iba07, koc08}. 
%%%%%%%%%%%%%%%%%%%%%%%%%%%%%%%%%%%%%%%%%%%%%%%%%%%%%%%%%%%%%%%%%%%%%%%%%%%%%%

%%%%%%%%%%%%%%%%%%%%%%%%%%%%%%%%%%%%%%%%%%%%%%%%%%%%%%%%%%%%%%%%%%%%%%%%%%%%%%
Recent observations of red giant stars near M31 have revealed a giant stellar stream to its south as well as giant stellar shells to the east and west of its center \citep{iba01, fer02, mcc03, iba04, iba05, guh06, koc08}. 
The giant stellar stream extends out to over $100\,\mathrm{kpc}$ from M31's center \citep{mcc03}. 
$N$-body simulations of the interaction between the progenitor of the giant stellar stream and M31 \citep{far07, far12, mor08} suggest that the stream, northeast shell, and west shell are tidal debris formed during the last pericentric passage of a satellite on a radial orbit. 
%%%%%%%%%%%%%%%%%%%%%%%%%%%%%%%%%%%%%%%%%%%%%%%%%%%%%%%%%%%%%%%%%%%%%%%%%%%%%%

%%%%%%%%%%%%%%%%%%%%%%%%%%%%%%%%%%%%%%%%%%%%%%%%%%%%%%%%%%%%%%%%%%%%%%%%%%%%%%
After the first reproduction of the giant stellar stream using $N$-body simulation by \citet{far07}, many studies based on $N$-body simulations have devoted to investigating various aspects of the observed structures. 
\citet{mor08} investigated the dynamical response of the M31 disk in detail and derived a mass range of the progenitor dwarf galaxy. 
\citet{far08} and \citet{sad13} showed a collision model of a disk galaxy with M31 also reproduces the observed structures well. 
\citet{far13} improved the collision model of \citet{far07} in various aspects (e.g., the infalling orbit of the progenitor and the mass of M31). 
\citet{ham10, ham13} proposed an alternative scenario that a past major merger produces M31, the giant stellar stream and stellar shells. 
The results of the minor merger scenario based on \citet{far07} have been compared with results of spectroscopic observations. 
Observations by \citet{gil07, gil09, koc08} discovered additional structures on phase space predicted by \citet{far07}. 
\citet{far12} reported a beautiful agreement of their observation and $N$-body simulation in the west shell region. 
\citet{kir14} investigated the density profile of the dark matter halo in M31 using an $N$-body simulation. 
To reproduce the giant stellar stream and the stellar shells, the density profile of the dark matter halo in M31 must be steeper than that of the prediction of the cold dark matter model. 
\citet{mik14} and \citet{kaw14} predicted that a wandering supermassive black hole lies within the halo (20--50 kpc from the M31 center). 
%%%%%%%%%%%%%%%%%%%%%%%%%%%%%%%%%%%%%%%%%%%%%%%%%%%%%%%%%%%%%%%%%%%%%%%%%%%%%%

%%%%%%%%%%%%%%%%%%%%%%%%%%%%%%%%%%%%%%%%%%%%%%%%%%%%%%%%%%%%%%%%%%%%%%%%%%%%%%
Photometric observations represented by PAndAS \citep[Pan-Andromeda Archaeological Survey:][]{mcc09, ric11, mar13} discovered a few tens of satellite galaxies around M31. 
Recent spectroscopic observations such as \citet{col13} and SPLASH Survey \citep[Spectroscopic and Photometric Landscape of Andromeda's Stellar Halo:][]{kal10, tol12} obtained kinematic information on newly discovered dwarf spheroidal galaxies. 
The above-mentioned results of recent observations strongly accelerate investigation for physical properties of M31 dwarf spheroidal galaxies. 
As another strategy, investigating physical properties of the progenitor dwarf galaxy is also possible by comparing the observed structures with results of $N$-body simulation of a galaxy collision with M31. 
Information related to dynamics of the progenitor dwarf galaxy would be conserved as footprints in the observed structures. 
Destructive tests utilizing $N$-body simulation have the potential to recover fossil information on dynamics of the progenitor dwarf galaxy imprinted in the observed structures. 
%%%%%%%%%%%%%%%%%%%%%%%%%%%%%%%%%%%%%%%%%%%%%%%%%%%%%%%%%%%%%%%%%%%%%%%%%%%%%%

%%%%%%%%%%%%%%%%%%%%%%%%%%%%%%%%%%%%%%%%%%%%%%%%%%%%%%%%%%%%%%%%%%%%%%%%%%%%%%
Recent photometric observations \citep{iba07, mcc09, mar13, lew13} discovered many stellar structures (Streams A, B, C, and D, North West stream, South West Cloud, and the Eastern Cloud) in the M31 halo in addition to the giant stellar stream and the east and the west shell. 
So far, \citet{far08} demonstrated that some of the streams arise from the progenitor of the giant stellar stream. 
On the other hand, \citet{con16} showed that Streams C or D does not have the same origin with the giant stellar stream by measuring heliocentric distances to Streams C and D and the giant stellar stream. 
Thus, origins of these structures and their relations are not yet understood, and are still open questions. 
In this paper, we focus on the giant stellar stream and its progenitor by assuming the recently discovered structures have origins distinct from the giant stream. 
%%%%%%%%%%%%%%%%%%%%%%%%%%%%%%%%%%%%%%%%%%%%%%%%%%%%%%%%%%%%%%%%%%%%%%%%%%%%%%

%%%%%%%%%%%%%%%%%%%%%%%%%%%%%%%%%%%%%%%%%%%%%%%%%%%%%%%%%%%%%%%%%%%%%%%%%%%%%%
The remainder of this paper is organized as follows. 
In \S\ref{section:model}, we describe the M31 model, including the disk, bulge, and dark matter halo and the satellite models. 
In \S\ref{section:results}, we present the results of the numerical simulations and analyze them. 
Finally, in \S\ref{section:discussion}, we summarize results and compare the results with observations. 
%%%%%%%%%%%%%%%%%%%%%%%%%%%%%%%%%%%%%%%%%%%%%%%%%%%%%%%%%%%%%%%%%%%%%%%%%%%%%%
%%%%%%%%%%%%%%%%%%%%%%%%%%%%%%%%%%%%%%%%%%%%%%%%%%%%%%%%%%%%%%%%%%%%%%%%%%%%%%

%%%%%%%%%%%%%%%%%%%%%%%%%%%%%%%%%%%%%%%%%%%%%%%%%%%%%%%%%%%%%%%%%%%%%%%%%%%%%%
%%%%%%%%%%%%%%%%%%%%%%%%%%%%%%%%%%%%%%%%%%%%%%%%%%%%%%%%%%%%%%%%%%%%%%%%%%%%%%
\section{Model description of interaction between M31 and satellite galaxies}
\label{section:model}
%%%%%%%%%%%%%%%%%%%%%%%%%%%%%%%%%%%%%%%%%%%%%%%%%%%%%%%%%%%%%%%%%%%%%%%%%%%%%%
%%%%%%%%%%%%%%%%%%%%%%%%%%%%%%%%%%%%%%%%%%%%%%%%%%%%%%%%%%%%%%%%%%%%%%%%%%%%%%
\begin{deluxetable*}{ccllllllll}
  \tabletypesize{\scriptsize}
  \tablewidth{0pt}
  \tablecaption{Parameters of fiducial models \label{tab.model}}
  \tablehead{ 
    \colhead{model/panel} 
    & \colhead{$M_{\rm sat} (M_{\odot})$} 
    & \colhead{$c$} 
    & \colhead{$W_{\rm 0}$\tablenotemark{(1)}} 
    & \colhead{$r_{\rm t}$ (kpc)} 
    & \colhead{$r_{\rm 0}$ (kpc)} 
    & \colhead{$\sigma_{\rm 0}$ (km s$^{-1}$)\tablenotemark{(2)}} %% one-dimensional velocity dispersion at the center
    & \colhead{$\rho_{\rm 0}$ ($M_{\odot}$ pc$^{-3}$)\tablenotemark{(3)}} 
    & \colhead{$\Sigma_{\rm 0}$ ($M_{\odot}$ pc$^{-2}$)\tablenotemark{(4)}} 
    & \colhead{$\Phi_{\rm r0}$ (erg g$^{-1}$)\tablenotemark{(5)}} 
  }
  \startdata
  A & $3\times10^9$ & 0.7 & 3.0  & 4.5 & 0.96 &  49.1 & $6.6\times10^{-1}$ & $9.6\times10^2$ & $-1.0\times10^{14}$ \\
  B & $1\times10^9$ & 0.5 & 1.9  & 2.0 & 0.65 &  40.7 & $1.4\times10^0$   & $1.1\times10^3$ & $-5.8\times10^{13}$ \\
  C & $1\times10^9$ & 0.1 & 0.44 & 6.0 & 4.90 &  22.6 & $2.9\times10^{-2}$ & $9.1\times10^1$ & $-8.8\times10^{12}$ \\
  D & $2\times10^9$ & 1.1 & 5.3  & 1.5 & 0.12 &  86.2 & $9.4\times10^1$   & $2.1\times10^4$ & $-4.2\times10^{14}$ \\
  E & $2\times10^9$ & 1.1 & 5.3  & 0.5 & 0.04 & 149.3 & $2.5\times10^3$   & $1.9\times10^5$ & $-1.2\times10^{15}$
  \enddata
  \tablenotetext{(1)}{Dimensionless King parameter at the center of the satellite. }
  \tablenotetext{(2)}{One-dimensional velocity dispersion at the center of the satellite. }
  \tablenotetext{(3)}{Mass density at the center of the satellite. }
  \tablenotetext{(4)}{Column mass density at the center of the satellite. }
  \tablenotetext{(5)}{Potential at the core radius $r_{\rm 0}$. }
\end{deluxetable*}
%%%%%%%%%%%%%%%%%%%%%%%%%%%%%%%%%%%%%%%%%%%%%%%%%%%%%%%%%%%%%%%%%%%%%%%%%%%%%%
We performed 247 comprehensive $N$-body simulations of the interaction between M31 and the progenitor dwarf galaxy with different sizes and density profiles to explore the characteristics of the progenitor of the giant stream. 
We represented the progenitor dwarf using King spheres and modeled the gravitational potential of M31 by a fixed potential, as described below in \S\ref{subsec:M31}. 
%%%%%%%%%%%%%%%%%%%%%%%%%%%%%%%%%%%%%%%%%%%%%%%%%%%%%%%%%%%%%%%%%%%%%%%%%%%%%%

%%%%%%%%%%%%%%%%%%%%%%%%%%%%%%%%%%%%%%%%%%%%%%%%%%%%%%%%%%%%%%%%%%%%%%%%%%%%%%
\subsection{Model of M31}
\label{subsec:M31}
%%%%%%%%%%%%%%%%%%%%%%%%%%%%%%%%%%%%%%%%%%%%%%%%%%%%%%%%%%%%%%%%%%%%%%%%%%%%%%
To investigate the dynamical response of the orbiting satellite, we modeled the dwarf galaxy by a self-consistent $N$-body realization of stars under the influence of an external force provided by M31. 
For simplicity, we assume that M31 is composed of three components: a disk, a bulge, and a dark matter halo. 
It should be noted that \citet{mor08} studied the self-gravitating response of the disk, bulge, and dark matter halo of M31 to an accreting satellite and concluded that satellites less massive than $5\times10^9M_\odot$ had a negligible effect on the gravitational potential of M31. 
Consequently, in this study, we treated M31 as the source of a fixed gravitational potential. 
%%%%%%%%%%%%%%%%%%%%%%%%%%%%%%%%%%%%%%%%%%%%%%%%%%%%%%%%%%%%%%%%%%%%%%%%%%%%%%

%%%%%%%%%%%%%%%%%%%%%%%%%%%%%%%%%%%%%%%%%%%%%%%%%%%%%%%%%%%%%%%%%%%%%%%%%%%%%%
We model the bulge of M31 as a spherically symmetric mass distribution represented by a Hernquist profile \citep{her90}. 
The corresponding density-potential pair is given by
%%%%%%%%%%%%%%%%%%%%%%%%%%%%%%%%%%%%%%%%%%%%%%%%%%%%%%%%%%%%%%%%%%%%%%%%%%%%%%
\begin{eqnarray}
  \rho_{\rm b}(r) & = & \left( \frac{M_{\rm b}}{2\pi r_{\rm b}^3} \right) \frac{1}{(r/r_{\rm b})(1+r/r_{\rm b})^3}, \\
  \Phi_{\rm b}(r) & = &- \frac{G M_{\rm b}}{r_{\rm b}+r}, 
\end{eqnarray}
%%%%%%%%%%%%%%%%%%%%%%%%%%%%%%%%%%%%%%%%%%%%%%%%%%%%%%%%%%%%%%%%%%%%%%%%%%%%%%
where $M_{\rm b}=3.24\times 10^{10} M_{\odot}$ is the total mass of the bulge; $r_{\rm b}=0.61 \,{\rm kpc}$, its scale radius; and $G$, the gravitational constant. 
The density-potential pair of the axisymmetric distribution in cylindrical coordinates $(R,z)$ of the disk is given by 
%%%%%%%%%%%%%%%%%%%%%%%%%%%%%%%%%%%%%%%%%%%%%%%%%%%%%%%%%%%%%%%%%%%%%%%%%%%%%%
\begin{eqnarray}
  \rho_{\rm d}(R,z) & = & \frac{\Sigma_{\rm 0}}{2z_{\rm d}} \exp \left( -\frac{R}{R_{\rm d}} \right) \exp \left( -\frac{|z|}{z_{\rm d}} \right),\\
  \Phi_{\rm d}(R,z) & = &- \frac{2G \Sigma_{\rm 0}}{R_{\rm d}z_{\rm d}} 
  \int_{-\infty}^{\infty} dz^{\prime} \exp \left(-\frac{|z^{\prime}|}{z_{\rm d}} \right) \nonumber \\
  &&\int_{0}^{\infty} da \sin^{-1} \left( \frac{2a}{\sqrt{+}+\sqrt{-}} \right) a K_0 \left(\frac{a}{R_{\rm d}}\right),
\end{eqnarray}
%%%%%%%%%%%%%%%%%%%%%%%%%%%%%%%%%%%%%%%%%%%%%%%%%%%%%%%%%%%%%%%%%%%%%%%%%%%%%%
where $r=\sqrt{R^2+z^2}, \sqrt{\pm}=\sqrt{(z-z^{\prime})^2+(a \pm R)^2}$, $\Sigma_0=2.0\times 10^8 M_\odot {\rm kpc}^{-2}$ is the central surface density of the disk, $R_{\rm d}=5.40 \,{\rm kpc}$ is the disk scale radius, $z_{\rm d}=0.60 \,{\rm kpc}$ is the disk scale height, and $K_{\alpha}(x)$ is the modified Bessel function \citep[cf.][]{bin08}. 
In this case, the total mass of the disk is $M_{\rm d}=3.66\times10^{10} M_\odot$. 
%%%%%%%%%%%%%%%%%%%%%%%%%%%%%%%%%%%%%%%%%%%%%%%%%%%%%%%%%%%%%%%%%%%%%%%%%%%%%%

%%%%%%%%%%%%%%%%%%%%%%%%%%%%%%%%%%%%%%%%%%%%%%%%%%%%%%%%%%%%%%%%%%%%%%%%%%%%%%
Finally, we assume that the extended dark matter halo can be adequately modeled as a spherically symmetric system. 
Using $N$-body simulations, \citet{nav97} pointed out that the central cusp of the dark matter halo can be approximated as $\rho(r) \propto r^{-1}$. 
On the other hand, \citet{fuk97} used a high-resolution $N$-body simulation to obtain a central cusp steeper than that in the aforementioned study \citep[see also][]{moo99}. 
Although the exact exponent of the density in the inner halo structure has been widely debated, the resulting structure of the dark matter halos depends on the number of particles used in the simulation. 
In our model, the density of the bulge component dominates the inner part of the galaxy, and therefore, this issue can be ignored. 
Here, we adopt the Navarro-Frenk-White profile, and the density-potential pair is given by
%%%%%%%%%%%%%%%%%%%%%%%%%%%%%%%%%%%%%%%%%%%%%%%%%%%%%%%%%%%%%%%%%%%%%%%%%%%%%%
\begin{eqnarray}
  \rho_{\rm h}(r) & = & \frac{\delta_{\rm c} \rho_{\rm c}}{(r/r_{\rm h})(1+r/r_{\rm h})^2},\\
  \Phi_{\rm h}(r) & = & - 4 \pi G \delta_{\rm c} \rho_{\rm c} r_{\rm h}^2 \left( \frac{r_{\rm h}}{r}\right)
  \ln \left( 1+\frac{r}{r_{\rm h}} \right),
\end{eqnarray}
%%%%%%%%%%%%%%%%%%%%%%%%%%%%%%%%%%%%%%%%%%%%%%%%%%%%%%%%%%%%%%%%%%%%%%%%%%%%%%
where $\delta_{\rm c}=4.41\times10^5$ is the characteristic density relative to the present-day critical density $\rho_{\rm c}=277.72 \, h^2 \,M_\odot {\rm kpc}^{-2}$, $h=0.71$ is the Hubble constant, and $r_{\rm h}=7.63 \,{\rm kpc}$ is the halo scale radius. 
The total mass of the dark matter halo is $M_{200}=8.8\times10^{11} M_\odot$ within the virial radius $R_{\rm 200}=195 \,{\rm kpc}$. 
The specific parameters used here were carefully determined in \citet{gee06} and \citet{far06, far07}. 
%%%%%%%%%%%%%%%%%%%%%%%%%%%%%%%%%%%%%%%%%%%%%%%%%%%%%%%%%%%%%%%%%%%%%%%%%%%%%%

%%%%%%%%%%%%%%%%%%%%%%%%%%%%%%%%%%%%%%%%%%%%%%%%%%%%%%%%%%%%%%%%%%%%%%%%%%%%%%
\subsection{Initial condition of satellite}
\label{subsec:mass.range}
%%%%%%%%%%%%%%%%%%%%%%%%%%%%%%%%%%%%%%%%%%%%%%%%%%%%%%%%%%%%%%%%%%%%%%%%%%%%%%
\begin{figure}
  \plotone{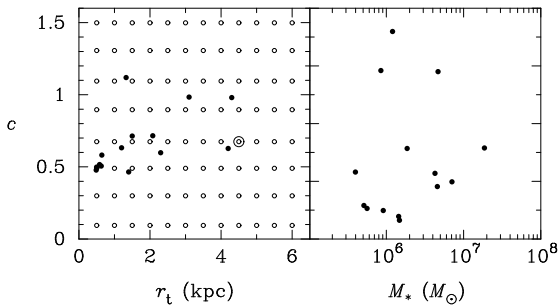}
  \caption{
    Relationship between the concentration $c$ and the tidal radius $r_{\mathrm{t}}$ (left panel) and the concentration $c$ and the stellar mass $M_{\rm *}$ (right panel) of local dwarf galaxies. 
    Data (filled circles) are compiled from \citet{woo08, mcc06, irw95}.
    The open circles represent satellite models tested in this study, and the double circle indicates the fiducial model (Model~A in Table~\ref{tab.model}). 
  }
  \label{local.dSPh}
\end{figure}
%%%%%%%%%%%%%%%%%%%%%%%%%%%%%%%%%%%%%%%%%%%%%%%%%%%%%%%%%%%%%%%%%%%%%%%%%%%%%%
Thus far, spherical progenitor models of the giant stream assumed a Plummer sphere with scale length of $1\,\mathrm{kpc}$ or a Hernquist sphere to represent the progenitor \citep{far07, far12, far13, mor08, sad13}. 
Because King profiles provide a tractable family of models with intuitive parameters that have been fitted extensively to nearby dwarf galaxies \citep{esk88a,esk88b,irw95,mcc06}, we employ King models with different sizes and density profiles to explore the characteristics of the progenitor of the giant stream.
Figure \ref{local.dSPh} shows the observed properties of the dwarf galaxies in the Local Group \citep{irw95,mcc06}, where $M_{\rm *}$ is the stellar mass; $r_{\mathrm{t}}$, the tidal radius; $c\equiv \log_{10}{r_{\mathrm{t}}/r_{\mathrm{0}}}$, the concentration parameter; and $r_{\mathrm{0}}$, the core radius of the King model. 
Based on these properties in the Local Group (Fig.~\ref{local.dSPh}) and the Virgo cluster \citep{ich86}, we performed a parameter study by varying the tidal radius from 0.5 to 6.0 kpc and the concentration parameter from 0.1 to 1.5.
%%%%%%%%%%%%%%%%%%%%%%%%%%%%%%%%%%%%%%%%%%%%%%%%%%%%%%%%%%%%%%%%%%%%%%%%%%%%%%

%%%%%%%%%%%%%%%%%%%%%%%%%%%%%%%%%%%%%%%%%%%%%%%%%%%%%%%%%%%%%%%%%%%%%%%%%%%%%%
To constrain the masses of the progenitor dwarfs of the giant stellar stream, \citet{mor08} estimated the disk heating as arising from the dynamical friction exerting a force opposite to the orbital motion. 
As a result, they found that the dynamical mass of the progenitor should be less than $5.2\times10^9 M_\odot$, because the disk thickness must agree with the observed thickness of M31 after the interaction of the satellite. 
Besides, the combination of the mass-metallicity relation of \citet{dek03} and the recent estimation of the heavy element abundance of the stream ${\rm [Fe/H]} \ga -1$ \citep{koc08} gives a lower mass limit of $5\times10^8 M_\odot$ for a progenitor stellar mass. 
Accordingly, the progenitor dwarfs most likely have a total mass in the range of $5\times10^8 M_\odot \la M_{\rm sat} \la 5\times 10^9 M_\odot$. 
Considering this estimation, we ran simulations for progenitor masses of $10^9 M_\odot$, $2\times10^9 M_\odot$, $3\times10^9 M_\odot$, and $5\times10^9 M_\odot$. 
%%%%%%%%%%%%%%%%%%%%%%%%%%%%%%%%%%%%%%%%%%%%%%%%%%%%%%%%%%%%%%%%%%%%%%%%%%%%%%

%%%%%%%%%%%%%%%%%%%%%%%%%%%%%%%%%%%%%%%%%%%%%%%%%%%%%%%%%%%%%%%%%%%%%%%%%%%%%%
\citet{far07} reported an $N$-body simulation of an accreting dwarf satellite within M31's fixed gravitational potential. 
They obtained orbital properties that are in good agreement with the observed properties of the giant stream. In addition, their simulation reproduced a photometric feature that they identified as the ``western shelf'' and ``northeast shelf.'' 
\citet{far12} presented a correspondence between the kinematics of the observed ``western shelf'' and that of the simulated one. 
\citet{mor08} also successfully reproduced these features using the same initial orbital elements of the progenitor in the case of a full self-gravitating system with a live disk, bulge, and dark matter halo. 
\citet{mik14} showed that the possible parameter space for the infalling orbit is limited to a very narrow region of the phase space. 
Moreover, they found that the possible parameter space includes Fardal's orbit. 
Therefore, we adopt Fardal's orbit in this study, and therefore, the initial position and velocity vector in the standard coordinates centered on M31 were $(-34.75$, $19.37$, $-13.99)\,\mathrm{kpc}$ and $(67.34$, $-26.12$, $13.50)\,\mathrm{km\,s^{-1}}$.
%%%%%%%%%%%%%%%%%%%%%%%%%%%%%%%%%%%%%%%%%%%%%%%%%%%%%%%%%%%%%%%%%%%%%%%%%%%%%%
%%%%%%%%%%%%%%%%%%%%%%%%%%%%%%%%%%%%%%%%%%%%%%%%%%%%%%%%%%%%%%%%%%%%%%%%%%%%%%

%%%%%%%%%%%%%%%%%%%%%%%%%%%%%%%%%%%%%%%%%%%%%%%%%%%%%%%%%%%%%%%%%%%%%%%%%%%%%%
%%%%%%%%%%%%%%%%%%%%%%%%%%%%%%%%%%%%%%%%%%%%%%%%%%%%%%%%%%%%%%%%%%%%%%%%%%%%%%
\section{Simulation results}
\label{section:results}
%%%%%%%%%%%%%%%%%%%%%%%%%%%%%%%%%%%%%%%%%%%%%%%%%%%%%%%%%%%%%%%%%%%%%%%%%%%%%%
\begin{figure*}
  \epsscale{1.2}
  %% \plotone{f2.eps}
  \plotone{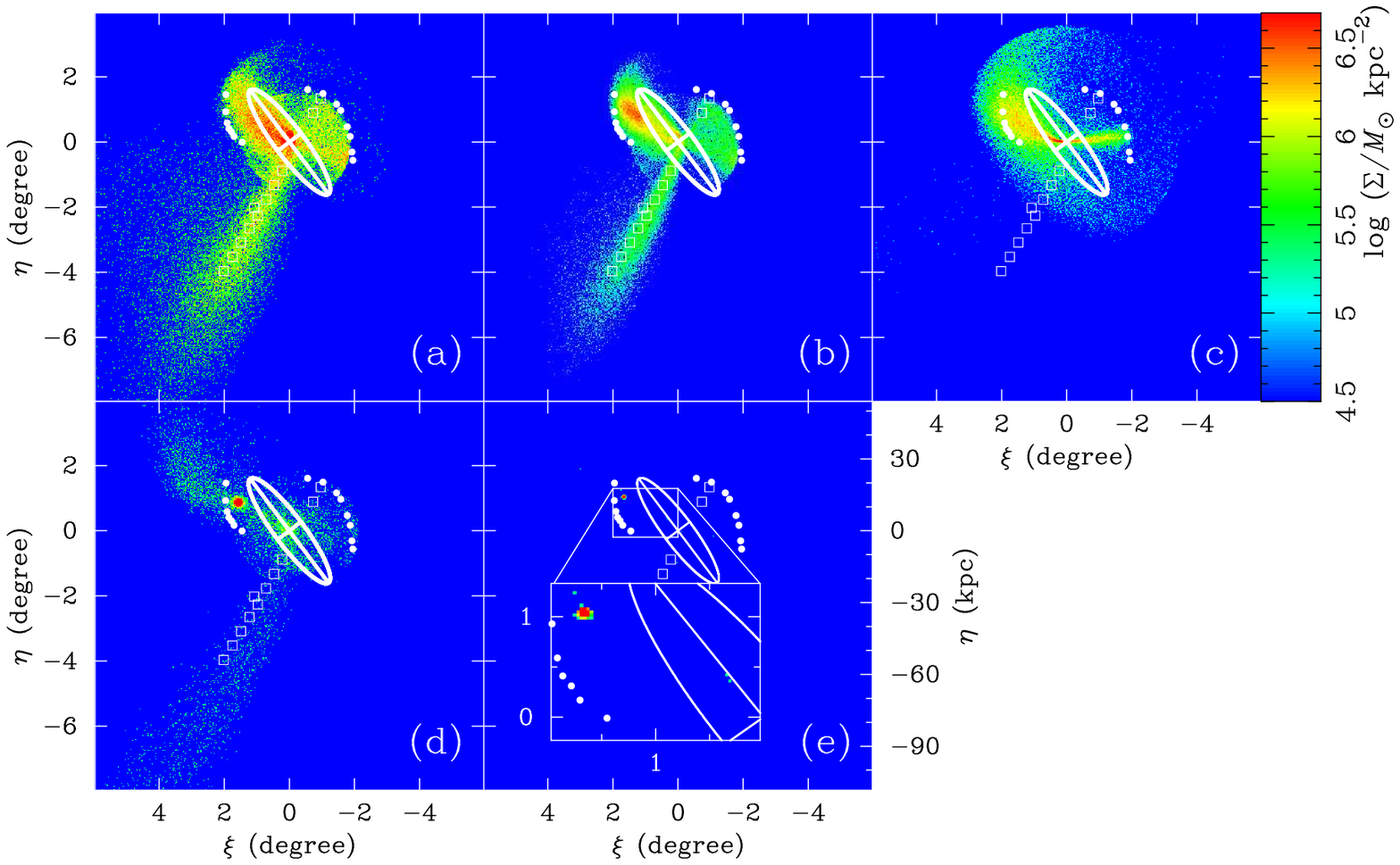}
  \caption{
    Projected mass-density distribution of the tidal debris for 
    (a) $M_{\rm sat}=3\times10^9 M_\odot$, $c=0.7$, $r_{\mathrm{t}}=4.5\,\mathrm{kpc}$, 
    (b) $M_{\rm sat}=1\times10^9 M_\odot$, $c=0.5$, $r_{\mathrm{t}}=2.0\,\mathrm{kpc}$, 
    (c) $M_{\rm sat}=1\times10^9 M_\odot$, $c=0.1$, $r_{\mathrm{t}}=6.0\,\mathrm{kpc}$, 
    (d) $M_{\rm sat}=2\times10^9 M_\odot$, $c=1.1$, $r_{\mathrm{t}}=1.5\,\mathrm{kpc}$, and 
    (e) $M_{\rm sat}=2\times10^9 M_\odot$, $c=1.1$, $r_{\mathrm{t}}=0.5\,\mathrm{kpc}$, respectively. 
    Filled circles and filled squares show the position of the edge of shells \citep{far07} and the observed areas of the Andromeda Giant Stellar Stream \citep{fon06}, respectively. 
    The ellipse in each panel corresponds the size of the M31's disk.
  }
  \label{fig2}
\end{figure*}
%%%%%%%%%%%%%%%%%%%%%%%%%%%%%%%%%%%%%%%%%%%%%%%%%%%%%%%%%%%%%%%%%%%%%%%%%%%%%%
\begin{figure*}
  \epsscale{1.2}
  %% \plotone{f3.eps}
  \plotone{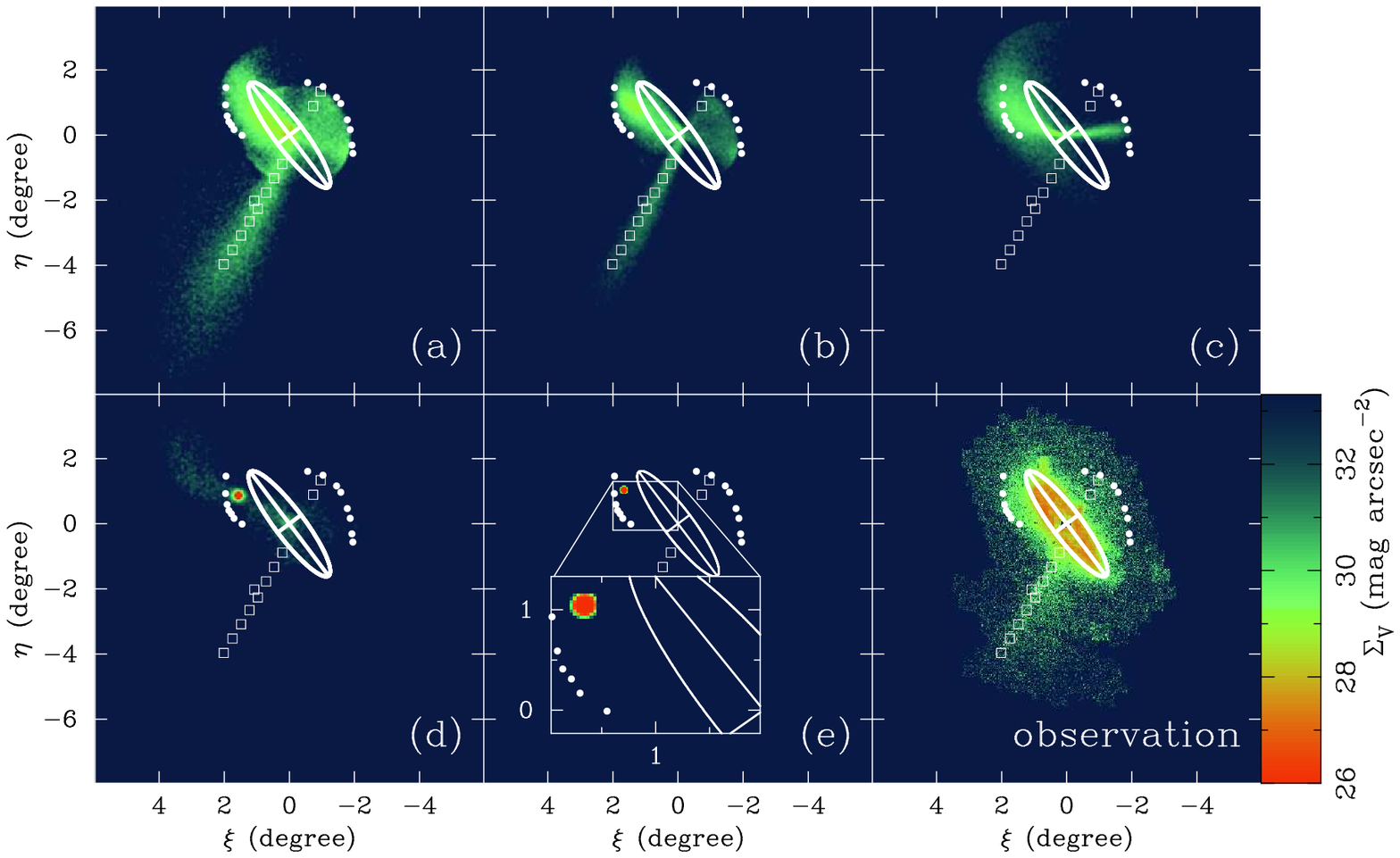}
  \caption{
    Surface brightness maps. 
    The lower-right panel shows a surface brightness map of the RGB stars around M31 observed by \citet{irw05}. 
    The other panels show $V$-band surface brightness map derived by our simulation results for 
    (a) $M_{\rm sat}=3\times10^9 M_\odot$, $c=0.7$, $r_{\mathrm{t}}=4.5\,\mathrm{kpc}$, 
    (b) $M_{\rm sat}=1\times10^9 M_\odot$, $c=0.5$, $r_{\mathrm{t}}=2.0\,\mathrm{kpc}$, 
    (c) $M_{\rm sat}=1\times10^9 M_\odot$, $c=0.1$, $r_{\mathrm{t}}=6.0\,\mathrm{kpc}$, 
    (d) $M_{\rm sat}=2\times10^9 M_\odot$, $c=1.1$, $r_{\mathrm{t}}=1.5\,\mathrm{kpc}$, and 
    (e) $M_{\rm sat}=2\times10^9 M_\odot$, $c=1.1$, $r_{\mathrm{t}}=0.5\,\mathrm{kpc}$, respectively. 
    Filled circles and filled squares show the position of the edge of shells \citep{far07} and the observed areas of the Andromeda Giant Stellar Stream \citep{fon06}, respectively. 
    The ellipse in each panel corresponds the size of the M31's disk.
  }
  \label{surface.brightness.map}
\end{figure*}
%%%%%%%%%%%%%%%%%%%%%%%%%%%%%%%%%%%%%%%%%%%%%%%%%%%%%%%%%%%%%%%%%%%%%%%%%%%%%%
We calculated 247 models in total: 49 for $M_{\rm sat}=10^9 M_\odot$, 87 for $M_{\rm sat}=2\times10^9 M_\odot$, 57 for $M_{\rm sat}=3\times10^9 M_\odot$, 53 for $M_{\rm sat}=5\times10^9 M_\odot$, and a Plummer model with total mass of $M_{\rm sat}=2\times10^9 M_\odot$ and effective radius of $1\,\mathrm{kpc}$ to check the consistency with \citet{far07}. 
In our simulation, we consider four mass models ($10^9 M_\odot$, $2\times10^9 M_\odot$, $3\times10^9 M_\odot$, and $5\times10^9 M_\odot$). 
For each mass model we vary the tidal radius and the concentration of the satellite, respectively, from 0.5~kpc to 6.0~kpc in 0.5~kpc increments (12 models in total) and 0.1 to 1.5 in 0.2 increments (8 models in total). 
This normally produces $4\times 12\times 8 = 384$ parameter sets; however, we have been able to reduce to number of models to run to 247. 
By running a coarse sampling of models, we were able to discover regions of the parameter space with $\chi^2_\nu > 3$ where the simulation fails to reproduce the observations; illustrated by deep-blue regions in Figure~\ref{contrast.c.rt.reduced.chi.square}. 
Because we densely sample a parameter space only where there is a plausible to match the observations, we are able to reduce to number of computationally sampled models to 247. 
Each model uses 65,536 particles to represent the King sphere, and the gravitational softening parameter is adopted as $\epsilon = r_{\mathrm{0}}/8$, which is sufficient to resolve the core radius $r_{\mathrm{0}}$. 
The direct $N$-body integration by the second-order Runge-Kutta method with an adaptive time step was performed on the FIRST simulator at the Center for Computational Sciences, University of Tsukuba. 
%%%%%%%%%%%%%%%%%%%%%%%%%%%%%%%%%%%%%%%%%%%%%%%%%%%%%%%%%%%%%%%%%%%%%%%%%%%%%%

%%%%%%%%%%%%%%%%%%%%%%%%%%%%%%%%%%%%%%%%%%%%%%%%%%%%%%%%%%%%%%%%%%%%%%%%%%%%%%
\subsection{Dynamical evolution of satellites}
\label{subsec:results:evolution}
%%%%%%%%%%%%%%%%%%%%%%%%%%%%%%%%%%%%%%%%%%%%%%%%%%%%%%%%%%%%%%%%%%%%%%%%%%%%%%
Figure~\ref{fig2} shows the projected particle positions for typical simulation results with different masses, tidal radii, and concentration parameters. 
Table~\ref{tab.model} lists the corresponding parameters for each panel in Fig.~\ref{fig2}. 
The ellipsoid in each panel indicates M31's disk size. 
The edge of the shells defined in \citet{far07} and the observed areas along the giant stream are indicated by filled circles and open squares, respectively. 
0.8 Gyr ago, the first pericentric passage close to the galactic center occurred, and the satellite collided almost head-on with the bulge. 
The distribution of satellite particles subsequently suffered tidal deformation and stretched out catastrophically in Models A, B, and C. 
In these models, this debris, while keeping a narrow distribution, expands to a great distance because a large fraction of the satellite particles acquire a high velocity relative to the center of M31. 
This creates the giant stream. After the second pericentric passage, stellar particles that initially constituted the satellite start to spread out in a fan-like form. 
A double shell system with roughly constant curvature is sharply defined, as seen in Figs.~\ref{fig2}a, \ref{fig2}b, and \ref{fig2}c. 
%%%%%%%%%%%%%%%%%%%%%%%%%%%%%%%%%%%%%%%%%%%%%%%%%%%%%%%%%%%%%%%%%%%%%%%%%%%%%%

%%%%%%%%%%%%%%%%%%%%%%%%%%%%%%%%%%%%%%%%%%%%%%%%%%%%%%%%%%%%%%%%%%%%%%%%%%%%%%
Models A and B successfully reproduce the stream and the shells at the east and west sides of M31, respectively. 
In contrast, Model C does not reproduce the observed structures very well because the stellar stream in Fig.~\ref{fig2}c is considerably shorter than the observed giant stream that extends out to over $100\,\mathrm{kpc}$ from the center of the M31 \citep{mcc03}. 
Furthermore, in this model, the shells have a narrower fan shape than do the observed ones, which have a large central angle. 
The central angle of the flagellum depends on the velocity dispersion of the progenitor satellite. 
That is, head-on collisions of the satellite with the shallower gravitational potential well generate the fan-like debris with the smaller central angle. 
Because the progenitor of Model C is a less massive fluffy galaxy with a larger core radius, it has a shallower gravitational potential and smaller velocity dispersion than those in Models A and B. 
Thus, the bunch of stars that are tidally stripped by M31's gravitational potential are not as spread out as in the observed fan-like structures. 
%%%%%%%%%%%%%%%%%%%%%%%%%%%%%%%%%%%%%%%%%%%%%%%%%%%%%%%%%%%%%%%%%%%%%%%%%%%%%%

%%%%%%%%%%%%%%%%%%%%%%%%%%%%%%%%%%%%%%%%%%%%%%%%%%%%%%%%%%%%%%%%%%%%%%%%%%%%%%
Figures \ref{fig2}d and \ref{fig2}e show the results of Models D and E, respectively. 
In both cases, the gravitational potential of the satellites is deeper than that in previous models because the progenitor has an appreciably small tidal radius. 
The distribution of satellite particles in Model D subsequently undergoes a tidal stripping after the first pericentric passage of the galactic center. 
The debris is drawn out into a long tail similar to the giant stream, but its stellar density is quite low. 
The stellar particles spread to form double shells in the same manner after the second pericentric passage. 
The shape of the shells is, however, quite different from the observed structures, and a high-density core of the progenitor still survives at $\xi \sim 2^{\circ}$ and $\eta \sim 1^{\circ}$, which is undetected by the observations. 
In an extreme case such as that in Model E, there are scarcely any tidal effects on such a compact satellite with the deep gravitational potential well. 
%%%%%%%%%%%%%%%%%%%%%%%%%%%%%%%%%%%%%%%%%%%%%%%%%%%%%%%%%%%%%%%%%%%%%%%%%%%%%%

%%%%%%%%%%%%%%%%%%%%%%%%%%%%%%%%%%%%%%%%%%%%%%%%%%%%%%%%%%%%%%%%%%%%%%%%%%%%%%
Surface brightness maps in $V$-band for the typical results are shown in Fig.~\ref{surface.brightness.map}. 
Here, we assume that the mass fraction of red giant branch (RGB) stars is 8\% using Salpeter's initial mass function. 
Then, we introduce a mass-to-light ratio for the observed flux so that the giant stream luminosity agrees with the observed luminosity $M_{\mathrm{V}}\approx-14$ \citep{iba01}. 
To compare the numerical results with observed results, the lower-right panel in Fig.~\ref{surface.brightness.map} shows the star count map by \citet{irw05}. 
Note that the PAndAS project \citep{mcc09, mar13} provides deeper observed data in the wider field; however, observed basic structures of the stream and the shells are virtually the same as \citet{irw05} and \citet{iba07}. 
Figures \ref{surface.brightness.map}a and \ref{surface.brightness.map}b nicely reproduce the observed features such as the Andromeda Stellar Stream and the shell-like structures. 
However, only Fig.~\ref{surface.brightness.map}a shows a clear third shell component. 
The other three panels show that these models failed to reproduce the observed features. 
%%%%%%%%%%%%%%%%%%%%%%%%%%%%%%%%%%%%%%%%%%%%%%%%%%%%%%%%%%%%%%%%%%%%%%%%%%%%%%

%%%%%%%%%%%%%%%%%%%%%%%%%%%%%%%%%%%%%%%%%%%%%%%%%%%%%%%%%%%%%%%%%%%%%%%%%%%%%%
\subsection{Mock images of simulated tidal debris}
\label{subsec:results:analysis}
%%%%%%%%%%%%%%%%%%%%%%%%%%%%%%%%%%%%%%%%%%%%%%%%%%%%%%%%%%%%%%%%%%%%%%%%%%%%%%
%%%%%%%%%%%%%%%%%%%%%%%%%%%%%%%%%%%%%%%%%%%%%%%%%%%%%%%%%%%%%%%%%%%%%%%%%%%%%%
\begin{figure}
  \plotone{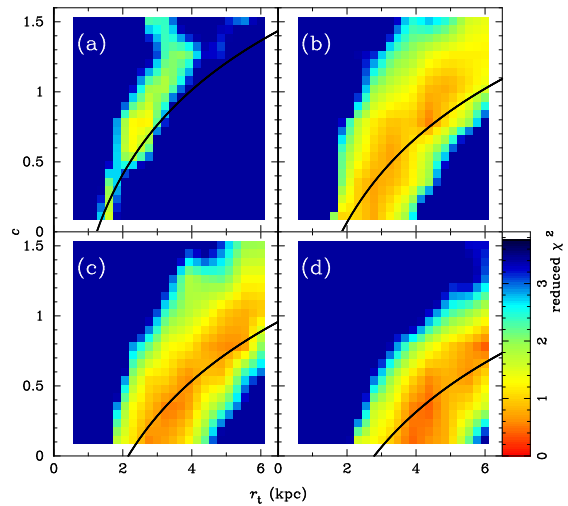}
  \caption{Reduced $\chi^2$ maps of the shapes of the both shells for 
    (a) $M_{\rm sat}=10^9 M_\odot$, 
    (b) $M_{\rm sat}=2\times10^9 M_\odot$, 
    (c) $M_{\rm sat}=3\times10^9 M_\odot$, and 
    (d) $M_{\rm sat}=5\times10^9 M_\odot$, respectively. 
    The horizontal axis is the tidal radius $r_{\mathrm{t}}$ of the progenitor dwarf galaxies and the vertical axis is the concentration parameter $c$. 
    Solid curve indicates the empirical relationship of the possible progenitor (see text).
  }
  \label{contrast.c.rt.reduced.chi.square}
\end{figure}
%%%%%%%%%%%%%%%%%%%%%%%%%%%%%%%%%%%%%%%%%%%%%%%%%%%%%%%%%%%%%%%%%%%%%%%%%%%%%%
In this subsection, we show qualitative comparisons between the simulation results and the observations. 
First, we test whether the velocity structure of the giant stellar stream is reproduced within an error range of $3\sigma$; the data were referred from Table 1 in \citet{fon06}, but we did not include the data of Field 8 because of the considerable contamination of the M31's disk components. 
Second, we checked the shapes of the northeast and west shells. 
The positions of each shell's edge were referred from Table 1 in \citet{far07}. 
The width of each shell's edge was estimated from the star count map in \citet{irw05}. 
Then, we obtained reduced $\chi^2$ maps of the shapes of the shells in the parameter space of $r_{\mathrm{t}}$ and $c$. 
In Fig.~\ref{contrast.c.rt.reduced.chi.square}, possible parameter regions of the progenitors are indicated by red or orange regions, which have a small, reduced $\chi^2$. 
``Possible'' regions are distributed from low $r_{\mathrm{t}}$ - low $c$ regions to high $r_{\mathrm{t}}$ - high $c$ regions. 
%%%%%%%%%%%%%%%%%%%%%%%%%%%%%%%%%%%%%%%%%%%%%%%%%%%%%%%%%%%%%%%%%%%%%%%%%%%%%%

%%%%%%%%%%%%%%%%%%%%%%%%%%%%%%%%%%%%%%%%%%%%%%%%%%%%%%%%%%%%%%%%%%%%%%%%%%%%%%
The interpretation of results in the previous subsection suggests that it is the depth of the potential well of the satellite --e.g., the binding gravitation-- that determines whether or not the simulation reproduces the observed structures. 
Here, we show this expectation explains the results well. 
First, the potential depends on the mass and the size of the satellite. 
The potential is given by $\Phi=-GM/R$, where $R$ is the typical radius ($r_{\rm 0}$ for the King model) and $M$ is the typical mass. 
From this relationship, $R$ should be proportional to $M$ to keep the potential constant. 
Second, the potential also depends on the mass distribution profile of the satellite. 
If $r_{\mathrm{t}}$ increases, the potential decreases. 
To conserve the potential of the central region (the most significant radius is the Hill radius: it determines whether stars are bound or stripped), the central density must increase. 
This implies that $r_{\mathrm{0}}$ must decrease against the increase of $r_{\mathrm{t}}$. 
Incorporating the two constraints, we find:
%%%%%%%%%%%%%%%%%%%%%%%%%%%%%%%%%%%%%%%%%%%%%%%%%%%%%%%%%%%%%%%%%%%%%%%%%%%%%%
\begin{equation}
  r_{\rm 0} = a \times M_{\rm sat} \times {r_{\rm t}}^{-1},
  \label{eq:fitting.relation0}
\end{equation}
%%%%%%%%%%%%%%%%%%%%%%%%%%%%%%%%%%%%%%%%%%%%%%%%%%%%%%%%%%%%%%%%%%%%%%%%%%%%%%
where $a$ is a remaining fitting parameter. 
By fitting $a$ for ``possible'' regions in Fig.~\ref{contrast.c.rt.reduced.chi.square}, an empirical relationship (the black curve in the figure) is derived as 
%%%%%%%%%%%%%%%%%%%%%%%%%%%%%%%%%%%%%%%%%%%%%%%%%%%%%%%%%%%%%%%%%%%%%%%%%%%%%%
\begin{equation}
  r_{\rm 0} = 1.0 \,{\rm kpc} \times \left(\frac{M_{\rm sat}}{3 \times 10^9 \, M_\odot}\right) \times \left(\frac{r_{\rm t}}{4.5\, {\rm kpc}}\right)^{-1}. 
  \label{eq:fitting.relation2}
\end{equation}
%%%%%%%%%%%%%%%%%%%%%%%%%%%%%%%%%%%%%%%%%%%%%%%%%%%%%%%%%%%%%%%%%%%%%%%%%%%%%%
The relationship agrees with the results of $N$-body simulations. 
Therefore, we conclude that the potential of the satellite is the key quantity to explain the dependence of ``possible'' regions on $M_{\rm sat}$, $r_{\rm t}$, and $c$. 
%%%%%%%%%%%%%%%%%%%%%%%%%%%%%%%%%%%%%%%%%%%%%%%%%%%%%%%%%%%%%%%%%%%%%%%%%%%%%%

%%%%%%%%%%%%%%%%%%%%%%%%%%%%%%%%%%%%%%%%%%%%%%%%%%%%%%%%%%%%%%%%%%%%%%%%%%%%%%
From the above discussion, we confirm that ``possible'' progenitors have the same general form of their potential well. 
If bound too strongly, the progenitors cannot be stripped to produce the stream and observed structures. 
Even if they are stripped, the numbers of stars are too small. 
If bound too weakly, then stripped stars cannot spread over a large enough volume to produce the observed structures. 
This is because the stellar velocity dispersion is too low in systems under the weak gravitational potential. 
Therefore, only progenitors with a suitable degree of binding potential are able to reproduce the observed structures. 
The relationship between the progenitor's mass and the area of the ``possible'' domain in the parameter space can be understood from this explanation. 
If the mass of a progenitor increases, then the gravitational potential increases. 
Therefore, the size of the progenitor must be larger to reproduce the observed structures. 
Therefore, the width of ``possible'' regions increases if the mass of the progenitor increases. 
%%%%%%%%%%%%%%%%%%%%%%%%%%%%%%%%%%%%%%%%%%%%%%%%%%%%%%%%%%%%%%%%%%%%%%%%%%%%%%
%%%%%%%%%%%%%%%%%%%%%%%%%%%%%%%%%%%%%%%%%%%%%%%%%%%%%%%%%%%%%%%%%%%%%%%%%%%%%%

%%%%%%%%%%%%%%%%%%%%%%%%%%%%%%%%%%%%%%%%%%%%%%%%%%%%%%%%%%%%%%%%%%%%%%%%%%%%%%
%%%%%%%%%%%%%%%%%%%%%%%%%%%%%%%%%%%%%%%%%%%%%%%%%%%%%%%%%%%%%%%%%%%%%%%%%%%%%%
\section{Summary and Discussion}
\label{section:discussion}
%%%%%%%%%%%%%%%%%%%%%%%%%%%%%%%%%%%%%%%%%%%%%%%%%%%%%%%%%%%%%%%%%%%%%%%%%%%%%%
We have studied the interaction between an accreting satellite dwarf galaxy and the Andromeda Galaxy using $N$-body simulations. 
A detailed parameter study is performed by varying the size and mass distribution of the progenitor dwarf galaxy. 
Our results showed that it is important to consider the strength of initial binding when reproducing the giant stellar stream and shells. 
%%%%%%%%%%%%%%%%%%%%%%%%%%%%%%%%%%%%%%%%%%%%%%%%%%%%%%%%%%%%%%%%%%%%%%%%%%%%%%

%%%%%%%%%%%%%%%%%%%%%%%%%%%%%%%%%%%%%%%%%%%%%%%%%%%%%%%%%%%%%%%%%%%%%%%%%%%%%%
In the following, we will discuss the implication of the nearby dwarf galaxies (\S\ref{subsec:nearby.dwarf}), possible existence of the third shell components depending on the properties of the progenitor (\S\ref{subsec:third.shell}), the bimodality of the radial velocity distribution in the giant stellar stream (\S\ref{subsec:bimodality}), and the inherent metallicity gradient of the progenitor satellite galaxies (\S\ref{subsec:metallicity}). 
%%%%%%%%%%%%%%%%%%%%%%%%%%%%%%%%%%%%%%%%%%%%%%%%%%%%%%%%%%%%%%%%%%%%%%%%%%%%%%
\subsection{Comparing the Impactor with Nearby Dwarf Galaxies}
\label{subsec:nearby.dwarf}
%%%%%%%%%%%%%%%%%%%%%%%%%%%%%%%%%%%%%%%%%%%%%%%%%%%%%%%%%%%%%%%%%%%%%%%%%%%%%%
\begin{figure}
  \centering
  %% \plotone{f5.eps}
  \plotone{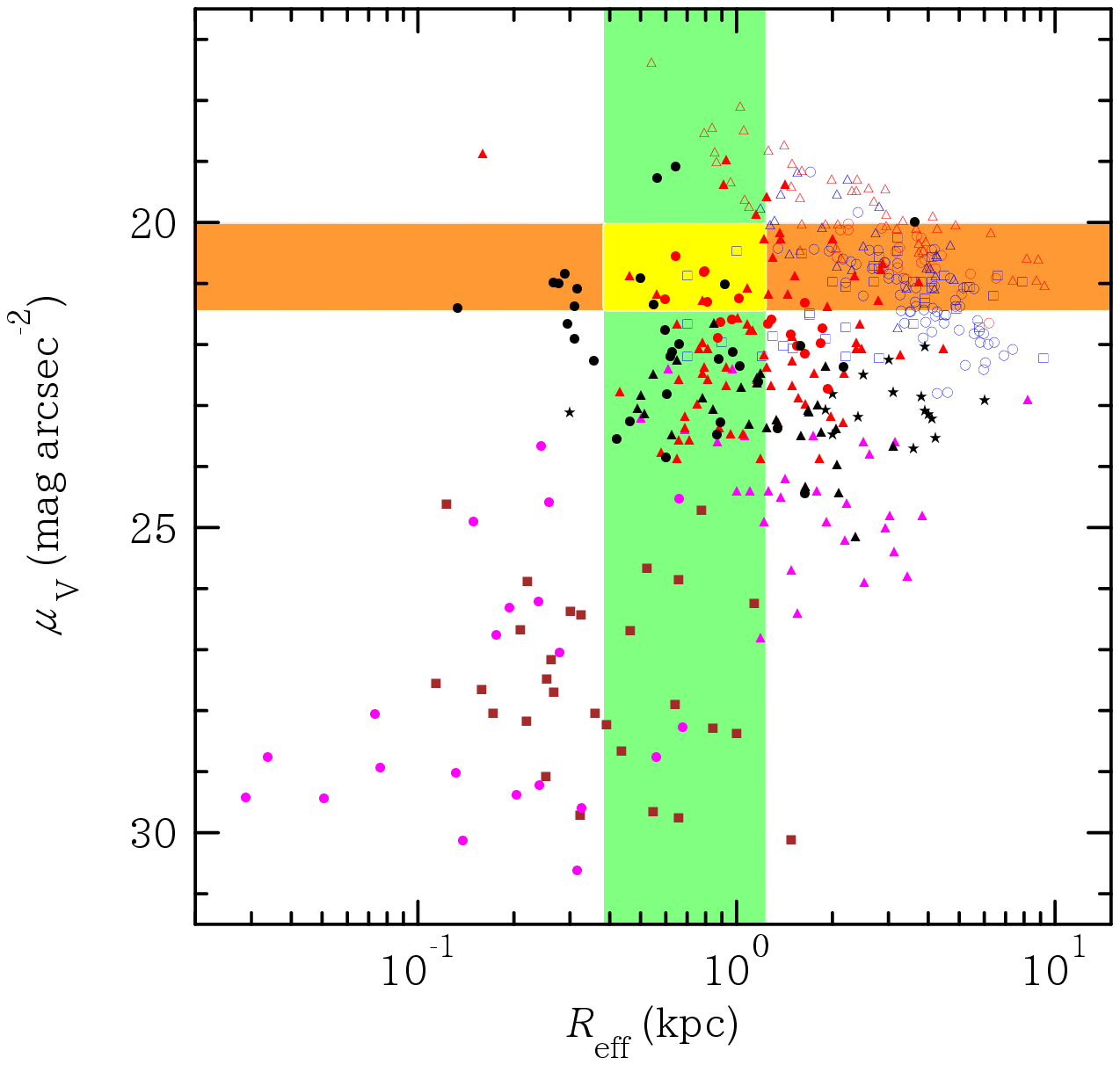}
  \caption{
    $V$-band surface brightness $\mu_{\rm V}$ as a function of effective radius $R_{\rm eff}$ (Kormendy relation in $V$-band). 
    Open symbols are the observed locations of galaxies: E/S0 galaxies (red circles for \citet{ken84} and red triangles for \citet{fal11}) and spiral galaxies (blue circles for \citet{ken84}, blue squares for \citet{kru87}, and blue triangles for \citet{fal11}). 
    Stars and magenta triangles show the properties of low surface brightness galaxies studied by \citet{blo95} and \citet{mat98}, respectively. 
    The remainder of the symbols show the observed properties for various types of dwarf galaxies: dwarf ellipticals in the Virgo cluster observed by \citet{tlb11, tlb12} (red filled circles), dwarf ellipticals in the Coma cluster observed by \citet{kou12a, kou12b} (red filled triangles), nearby dwarf galaxies observed by \citet{mak99} (black filled circles; most of them are dwarf irregulars), low luminosity dwarf irregulars in the Virgo cluster observed by \citet{hel01} (black filled triangles), dwarf spheroidals in the MW halo (magenta circles: \citet{bra11, wol10}), and all known Andromeda dwarf spheroidals compiled by \citet{col13} (brown squares). 
    The orange band show the empirical relation (Eq.~(\ref{eq:fitting.relation2}): solid curves in Fig.~\ref{contrast.c.rt.reduced.chi.square}) under an assumption of mass independent mass-to-light ratio with $1\sigma$ scatter of Faber-Jackson relation for model A \citep{tlb12, fal11}. 
    Green hatched region is the mass range for the progenitor dwarf galaxy derived by \citet{mor08}. 
    Parameter space within yellow hatched region shows physical properties of the possible progenitor dwarf galaxy. 
  }
  \label{fig:kormendy}
\end{figure}
%%%%%%%%%%%%%%%%%%%%%%%%%%%%%%%%%%%%%%%%%%%%%%%%%%%%%%%%%%%%%%%%%%%%%%%%%%%%%%

%%%%%%%%%%%%%%%%%%%%%%%%%%%%%%%%%%%%%%%%%%%%%%%%%%%%%%%%%%%%%%%%%%%%%%%%%%%%%%
We compare the empirical relation (Eq.~(\ref{eq:fitting.relation2})) with the observed properties of nearby galaxies in Fig.~\ref{fig:kormendy}. 
We plot $V$-band surface brightness $\mu_{\rm V}$ of nearby galaxies as a function of effective radius $R_{\rm eff}$ in Fig.~\ref{fig:kormendy}. 
An orange band in the horizontal direction shows the empirical relation (Eq.~(\ref{eq:fitting.relation2})) for $c = 0.7$ (corresponds to model A at $M_{\rm sat} = 3 \times 10^9 M_\odot$). 
The empirical relation itself implies that surface density of the ``possible'' progenitor galaxy in the central region is independent on mass to keep the strength of gravitational binding. 
In Fig.~\ref{fig:kormendy}, we assume a mass independent mass-to-light ratio model for simplicity: the width of the band corresponds to $1\sigma$ scatter of Faber-Jackson relation for Model A \citep{tlb12, fal11}. 
The orange band contains many observed galaxies; therefore, we can conclude that the empirical relation stays in realistic parameter region. 
On the other hand, \citet{mor08} derived the mass range of the progenitor dwarf galaxy as $5\times 10^8 M_\odot \leq M_{\rm sat} \leq 5\times 10^9 M_\odot$ (a green band in Fig.~\ref{fig:kormendy}). 
The yellow rectangle in Fig.~\ref{fig:kormendy} shows overlapped region of orange and green bands, which means ``possible'' parameter region for the progenitor dwarf galaxy of the observed structures. 
Fig.~\ref{fig:kormendy} clearly shows that some of the nearby dwarf galaxies have similar photometric properties with the progenitor dwarf galaxy of the observed structures. 
%%%%%%%%%%%%%%%%%%%%%%%%%%%%%%%%%%%%%%%%%%%%%%%%%%%%%%%%%%%%%%%%%%%%%%%%%%%%%%

%%%%%%%%%%%%%%%%%%%%%%%%%%%%%%%%%%%%%%%%%%%%%%%%%%%%%%%%%%%%%%%%%%%%%%%%%%%%%%
Here, we discuss the morphology of the progenitor dwarf galaxy.
The yellow rectangle in Fig.~\ref{fig:kormendy} contains not only dwarf elliptical galaxies but also dwarf irregulars and small spiral galaxies. 
We assume the spheroidal galaxy in this paper, however, the progenitor dwarf galaxy of the observed structures is possibly a dwarf irregular or dwarf spiral galaxy. 
Such a morphological difference can cause the different structures of the tidal debris after the collision with M31, and some of them might become a crucial point to solve current mismatches of $N$-body simulations with observations (for example, bimodality of the giant stream to be discussed in \S\ref{subsec:bimodality}). 
\citet{far08} and \citet{sad13} showed that infall model of a dwarf spiral galaxy toward M31 also reproduces the observed structures well. 
We will investigate the relationship between angular momentum of an infalling spiral galaxy and resultant structures and underlying physical mechanisms in a future study \citep{kir13}. 
%%%%%%%%%%%%%%%%%%%%%%%%%%%%%%%%%%%%%%%%%%%%%%%%%%%%%%%%%%%%%%%%%%%%%%%%%%%%%%

%%%%%%%%%%%%%%%%%%%%%%%%%%%%%%%%%%%%%%%%%%%%%%%%%%%%%%%%%%%%%%%%%%%%%%%%%%%%%%
\subsection{Third shell component}
\label{subsec:third.shell}
%%%%%%%%%%%%%%%%%%%%%%%%%%%%%%%%%%%%%%%%%%%%%%%%%%%%%%%%%%%%%%%%%%%%%%%%%%%%%%
\begin{figure}
  \plotone{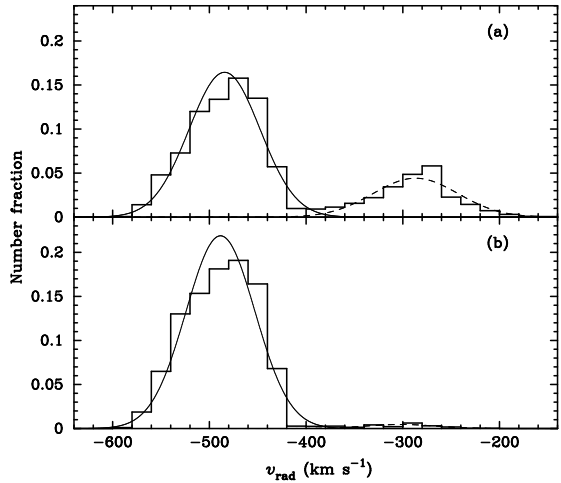}
  \caption{Histogram of the radial velocity around the field f207 \citep{gil09}. 
    In upper panel (Model A), there are two components: 75.6\% are $v_{\mathrm{rad}}=-487\pm 36\,\mathrm{km\,s^{-1}}$ (solid curve), 24.4\% are $v_{\mathrm{rad}}=-289\pm 44\,\mathrm{km\,s^{-1}}$ (dashed curve). 
    In bottom panel (Model B), there are also two components: 97.9\% are $v_{\mathrm{rad}}=-491\pm 35\,\mathrm{km\,s^{-1}}$ (solid curve), 2.1\% are $v_{\mathrm{rad}}=-306\pm 37\,\mathrm{km\,s^{-1}}$ (dashed curve). 
  }
  \label{third.on.off}
\end{figure}
%%%%%%%%%%%%%%%%%%%%%%%%%%%%%%%%%%%%%%%%%%%%%%%%%%%%%%%%%%%%%%%%%%%%%%%%%%%%%%
\citet{far07} pointed out that there is the third shell structure originating from the same progenitor in addition to the giant stellar stream, the northeast shell, and the west shell. 
A similar structure was also reported in \citet{mor08}. 
They showed that the third shell component is a forward continuation of the giant stellar stream. 
In our results of simulations, many parameter sets indicate the third shell component, and these explain the giant stellar stream, northeast shell, and west shell (see Fig.~\ref{surface.brightness.map}a: $M_{\rm sat}=3\times10^9M_{\odot}$, $c=0.7$, $r_{\rm t}=4.5\,\mathrm{kpc}$). 
However, some parameter sets also nicely reproduce the giant stellar stream, northeast shell, and west shell without the third shell (see Fig.~\ref{surface.brightness.map}b: $M_{\rm sat}=1\times10^9M_{\odot}$, $c=0.5$, $r_{\rm t}=2.0\,\mathrm{kpc}$). 
This is clearly different from earlier studies, and it suggests that the observed third shell component might not be a forward continuation of the giant stellar stream. 
It is important that both parameter sets explain the giant stellar stream, northeast shell, and west shell. 
%%%%%%%%%%%%%%%%%%%%%%%%%%%%%%%%%%%%%%%%%%%%%%%%%%%%%%%%%%%%%%%%%%%%%%%%%%%%%%

%%%%%%%%%%%%%%%%%%%%%%%%%%%%%%%%%%%%%%%%%%%%%%%%%%%%%%%%%%%%%%%%%%%%%%%%%%%%%%
To clarify our statement, we compared the velocity distribution of both cases. 
Figure \ref{third.on.off} shows the radial velocity histogram around field f207 (center of this region is $\xi=0\hspace{-.25mm}^{\circ}\hspace{-1mm}.2$, $\eta=-1\hspace{-.25mm}^{\circ}\hspace{-1mm}.3$). 
\citet{gil09} reported two more components exist besides the inner spheroid of M31: 31\% of the total population has $v_{\mathrm{rad}}=-524\pm 23\,\mathrm{km\,s^{-1}}$, and another 31\% has $v_{\mathrm{rad}}=-426\pm 21\,\mathrm{km\,s^{-1}}$. 
This figure can be compared with Fig.~6 in \citet{gil09}. The top panel shows a histogram of Model (a), and the bottom shows a Model (b) of Fig.~\ref{surface.brightness.map}. 
They are fitted by the Kayes mixture-modeling algorithm proposed in \citet{ash94}. 
The radial velocity of M31 is $-300\,\mathrm{km\,s^{-1}}$, with the negative sign implying that the direction of motion is toward us. 
Two components are clearly observed in (a): 75.6\% has $v_{\mathrm{rad}}=-487\pm 36\,\mathrm{km\,s^{-1}}$ (solid curve) and 24.4\% has $v_{\mathrm{rad}}=-289\pm 44\,\mathrm{km\,s^{-1}}$ (dashed curve). 
The former is a component of the giant stellar stream, and the latter is considered as the third shell component. 
The simulated result well explains the observation by \citet{gil09}, except a small difference of the contrast over the giant stream component. 
As shown in (b), there also exists the giant stellar stream component ($v_{\mathrm{rad}}=-491\pm 35\,\mathrm{km\,s^{-1}}$, solid curve) with a fraction of 97.9\%. 
The dashed curve in Fig.~\ref{third.on.off} is the kinematically hot component $v_{\mathrm{rad}}=-306\pm 37\,\mathrm{km\,s^{-1}}$; this is unlikely the third shell component. 
%%%%%%%%%%%%%%%%%%%%%%%%%%%%%%%%%%%%%%%%%%%%%%%%%%%%%%%%%%%%%%%%%%%%%%%%%%%%%%

%%%%%%%%%%%%%%%%%%%%%%%%%%%%%%%%%%%%%%%%%%%%%%%%%%%%%%%%%%%%%%%%%%%%%%%%%%%%%%
\citet{gil07,gil09} observed the velocity distribution of RGB stars in these regions to verify the existence of the third shell structure. 
The result shows there is more than one component: a giant stellar stream component and kinematically cold components. 
\citet{gil07} reported that some of new components are nearly consistent with the results of \citet{far07} (position, kinematic trends, and $\mathrm{[Fe/H]}$ distribution). 
However, the contrast between the two components does not match the stream component in the simulations: it in the simulations is much greater than that in the observations in each field of \citet{gil07}. 
Furthermore, even if the $\mathrm{[Fe/H]}$ distribution is consistent with the stream component, the two components could have different origins. 
More observations such as the $\mathrm{[Fe/Mg]}$ distribution will be required to determine whether the two components have the same or different origins, and it is important to analyse the simulation results in these regions precisely for future comparison. 
%%%%%%%%%%%%%%%%%%%%%%%%%%%%%%%%%%%%%%%%%%%%%%%%%%%%%%%%%%%%%%%%%%%%%%%%%%%%%%

%%%%%%%%%%%%%%%%%%%%%%%%%%%%%%%%%%%%%%%%%%%%%%%%%%%%%%%%%%%%%%%%%%%%%%%%%%%%%%
\subsection{Bimodality of giant stellar stream}
\label{subsec:bimodality}
%%%%%%%%%%%%%%%%%%%%%%%%%%%%%%%%%%%%%%%%%%%%%%%%%%%%%%%%%%%%%%%%%%%%%%%%%%%%%%
\begin{figure}
  \plotone{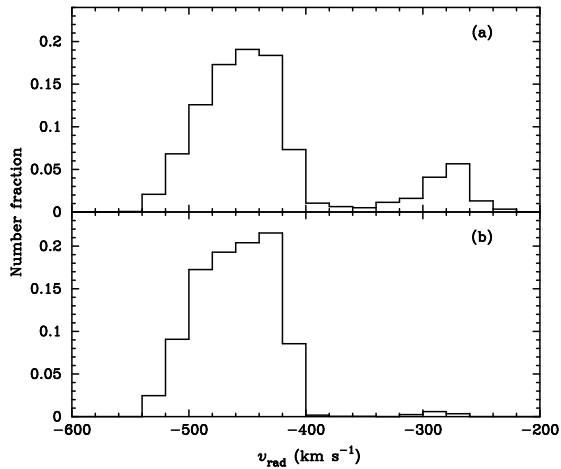}
  \caption{
    Histogram of the radial velocity around the field H13s \citep{koc08}. 
    Model (a) exhibits two clear components have radial velocity of $-460\pm 31\,\mathrm{km\,s^{-1}}$ and $-286\pm 24\,\mathrm{km\,s^{-1}}$ while model (b) has only one component of $v_{\mathrm{rad}}=-463\pm 30\,\mathrm{km\,s^{-1}}$. 
  }
  \label{H13s.region}
\end{figure}
%%%%%%%%%%%%%%%%%%%%%%%%%%%%%%%%%%%%%%%%%%%%%%%%%%%%%%%%%%%%%%%%%%%%%%%%%%%%%%
\begin{figure}
  \plotone{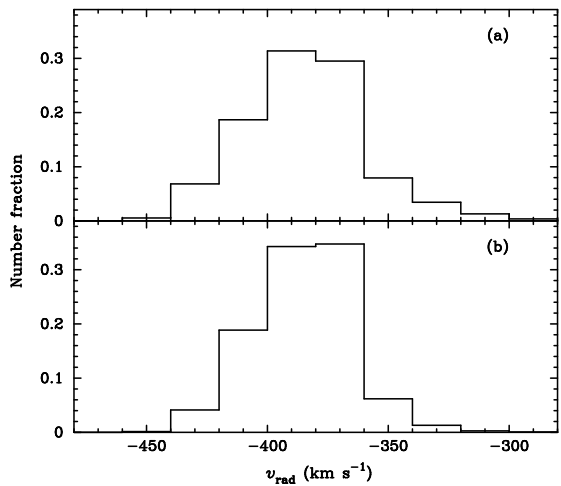}
  \caption{
    Histogram of the radial velocity around the field a3 \citep{koc08}. 
    Both the models (a) and (b) show a single component which has radial velocity around $-390\,\mathrm{km\,s^{-1}}$. 
  }
  \label{a3.region}
\end{figure}
%%%%%%%%%%%%%%%%%%%%%%%%%%%%%%%%%%%%%%%%%%%%%%%%%%%%%%%%%%%%%%%%%%%%%%%%%%%%%%

%%%%%%%%%%%%%%%%%%%%%%%%%%%%%%%%%%%%%%%%%%%%%%%%%%%%%%%%%%%%%%%%%%%%%%%%%%%%%%
The bimodality of the stream is observed in H13s region (center of this region is $\xi=0\hspace{-.25mm}^{\circ}\hspace{-1mm}.4$, $\eta=-1\hspace{-.25mm}^{\circ}\hspace{-1mm}.5$), and there are two components with peak radial velocity of $-520$ and $-400\,\mathrm{km\,s^{-1}}$ \citep{koc08}. 
\citet{gil09} performed more detailed analysis and reported radial velocity of two components are $-490\pm 21\,\mathrm{km\,s^{-1}}$ and $-388\pm 17\,\mathrm{km\,s^{-1}}$ for 48\% and 27\% of the total population, respectively. 
%%%%%%%%%%%%%%%%%%%%%%%%%%%%%%%%%%%%%%%%%%%%%%%%%%%%%%%%%%%%%%%%%%%%%%%%%%%%%%

%%%%%%%%%%%%%%%%%%%%%%%%%%%%%%%%%%%%%%%%%%%%%%%%%%%%%%%%%%%%%%%%%%%%%%%%%%%%%%
Figure \ref{H13s.region} shows the radial velocity histogram of our results around field H13s. 
In \citet{koc08}, both components cannot be considered as the third shell component. 
Therefore, we can neglect the peak radial velocity component of $\sim-290\,\mathrm{km\,s^{-1}}$ as the origin of bimodality. 
These figures show that there is no clear double component in the histogram except for the third shell. 
This is a natural result from the assumptions made in our simulations. 
We assume King models for the progenitor, and therefore, the progenitor is a single component in phase space. 
However, if the progenitor has two or more components in phase space, as is the case for dwarf spirals or dwarf irregulars, the results should be changed, and the observed bimodality might be reproduced. 
We will report it in the forthcoming study. 
From the other viewpoint, the observed bimodality might also be attributable to a different accretion event. 
This hypothesis could be supported by the analysis of field a3 (center of this region is $\xi=1\hspace{-.25mm}^{\circ}\hspace{-1mm}.3$, $\eta=-2\hspace{-.25mm}^{\circ}\hspace{-1mm}.1$) \citep{koc08}. 
The bimodality of the giant stellar stream is not observed in the a3 region \citep{koc08,gil09}, and our results also do not indicate bimodality in this region (Fig.~\ref{a3.region}). 
This result suggests that the H13s region is an unusual region and that bimodality is not a typical feature of the giant stellar stream. 
From the RGB star count map in \citet{irw05}, we can identify many structures except for the giant stellar stream, northeast shell, and west shell \citep[see also][]{mcc09,mar13}. 
There is a faint shell-like structure near field H13s. 
Detailed investigation of formation processes and the relationship to the giant stream of other structures (e.g., Streams A, B, C, and D) might indicate the necessary direction to go in reproducing the stream bimodality. 
%%%%%%%%%%%%%%%%%%%%%%%%%%%%%%%%%%%%%%%%%%%%%%%%%%%%%%%%%%%%%%%%%%%%%%%%%%%%%%

%%%%%%%%%%%%%%%%%%%%%%%%%%%%%%%%%%%%%%%%%%%%%%%%%%%%%%%%%%%%%%%%%%%%%%%%%%%%%%
More precise observations of the radial velocity, metallicity distribution or abundance pattern near field H13s, and other stream regions may shed additional light on the origin of the two observed components. 
%%%%%%%%%%%%%%%%%%%%%%%%%%%%%%%%%%%%%%%%%%%%%%%%%%%%%%%%%%%%%%%%%%%%%%%%%%%%%%

%%%%%%%%%%%%%%%%%%%%%%%%%%%%%%%%%%%%%%%%%%%%%%%%%%%%%%%%%%%%%%%%%%%%%%%%%%%%%%
\subsection{Metallicity Gradient of the Progenitor Satellite}
\label{subsec:metallicity}
%%%%%%%%%%%%%%%%%%%%%%%%%%%%%%%%%%%%%%%%%%%%%%%%%%%%%%%%%%%%%%%%%%%%%%%%%%%%%%
\begin{figure*}
  \centering
  %% %% \epsscale{0.94}
  %% \plotone{f9.eps}
  \plotone{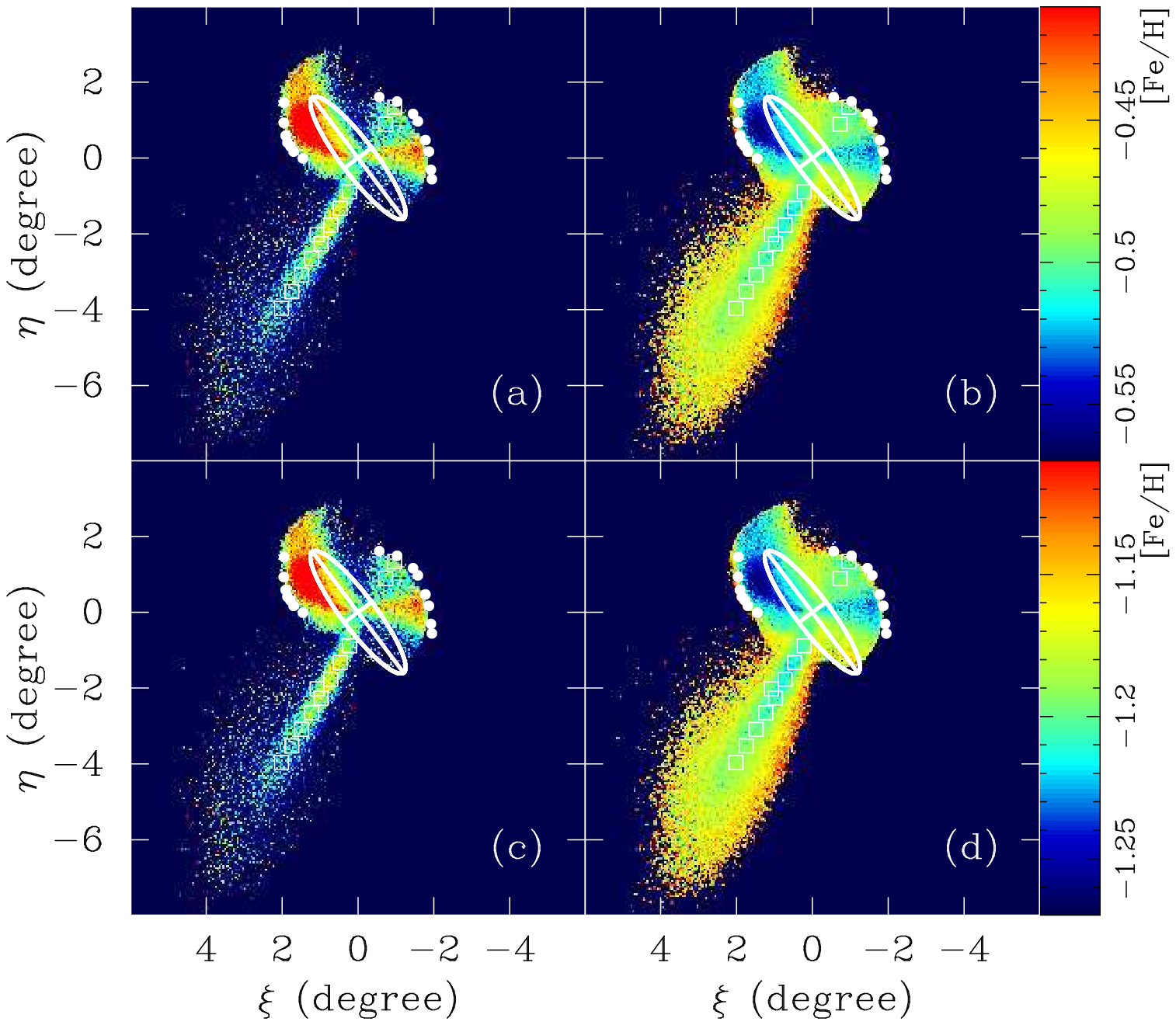}
  \caption{
    Metallicity distribution for 
    (a) ${\rm [Fe/H]}_{\rm mean} = -0.5$, $\Delta {\rm [Fe/H]} = -0.7$, 
    (b) ${\rm [Fe/H]}_{\rm mean} = -0.5$, $\Delta {\rm [Fe/H]} = 0.3$, 
    (c) ${\rm [Fe/H]}_{\rm mean} = -1.2$, $\Delta {\rm [Fe/H]} = -0.7$, 
    and 
    (d) ${\rm [Fe/H]}_{\rm mean} = -1.2$, $\Delta {\rm [Fe/H]} = 0.3$, 
    respectively. 
    Filled circles and open squares show the position of the edge of shells \citep{far07} and the observed areas of the giant stellar stream \citep{fon06}, respectively. 
    Ellipse in each panel corresponds the size of the M31's disk. 
  }
  \label{fig:stream:convergence:metallicity:map}
\end{figure*}
%%%%%%%%%%%%%%%%%%%%%%%%%%%%%%%%%%%%%%%%%%%%%%%%%%%%%%%%%%%%%%%%%%%%%%%%%%%%%%
\begin{figure}
  \centering
  \plotone{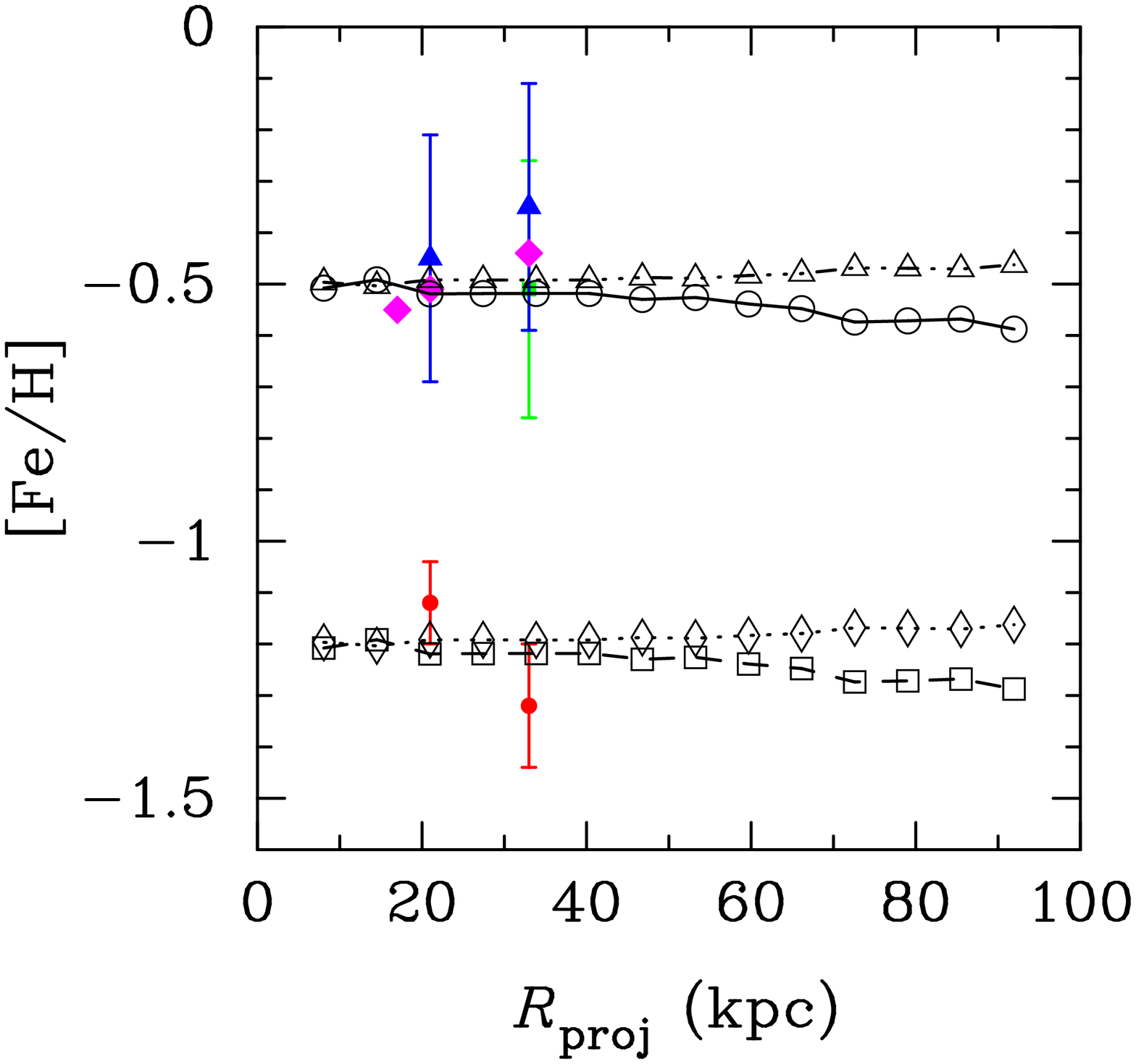}
  \caption{
    Radial profile of metallicity along the giant stellar stream. 
    Open symbols with lines represent metallicity distribution models for the result of $N$-body simulations shown in Fig.~\ref{fig:stream:convergence:metallicity:map}: 
    (a) by circles with a solid line, (b) by triangles with a triple-dot-dashed line, (c) by squares with a dashed line, and (d) by diamonds with a dotted line. 
    Filled symbols show the observed metallicity: red circles \citep{koc08}, green square \citep{guh06}, blue triangles \citep{Kalirai2006inner, Kalirai2006outer}, and magenta diamonds \citep{gil09}. 
  }
  \label{fig:stream:convergence:metallicity:profile}
\end{figure}
%%%%%%%%%%%%%%%%%%%%%%%%%%%%%%%%%%%%%%%%%%%%%%%%%%%%%%%%%%%%%%%%%%%%%%%%%%%%%%
The mass-metallicity relation and the metallicity gradient in a progenitor galaxy are important clues to the origin and nature of the progenitor. 
In this study, we define the metallicity gradient of a galaxy $\Delta {\rm [Fe/H]}$ as $d{\rm [Fe/H]}(r) / d\log{(r/r_{\rm e})}$, where ${\rm [Fe/H]}(r)$ is the radial profile of metallicity ${\rm [Fe/H]}$ and $r_{\rm e}$ is the effective radius. 
Recent observational studies reveal that the dwarf elliptical galaxies in the local universe exhibit gradients of either sign \citep[$-0.6 \lesssim \Delta {\rm [Fe/H]} \lesssim 0.2$ from][]{Spolaor2009,Koleva2009observation,Koleva2009summary}. 
However, the origin is still unclear \citep[e.g.][]{Mori1997,Mori1999}. 
%%%%%%%%%%%%%%%%%%%%%%%%%%%%%%%%%%%%%%%%%%%%%%%%%%%%%%%%%%%%%%%%%%%%%%%%%%%%%%

%%%%%%%%%%%%%%%%%%%%%%%%%%%%%%%%%%%%%%%%%%%%%%%%%%%%%%%%%%%%%%%%%%%%%%%%%%%%%
Studies focused on merger remnants would be a useful tool to explore the metallicity gradient of the corresponding progenitor. 
If a satellite galaxy initially had some non-uniform metallicity distribution, then structures formed after a galactic merger should also have a non-uniform metallicity distribution. 
Therefore, theoretical studies based on $N$-body simulations have the potential to connect the intrinsic metallicity distribution of the progenitor galaxy and the current metallicity distribution of merger remnants. 
\citet{far08} studied connections between satellite galaxy models which initially have a negative metallicity gradient and the resultant metallicity distribution of the giant stellar stream, the east and the west shells. 
However, there has been to date no comparison of the outcome for satellite models for positive versus negative gradients. 
It is difficult to determine the mean metallicity of the progenitor galaxy without the knowledge about the relationship between the intrinsic metal gradient of the progenitor galaxy and the observed metallicity distribution of the merger remnants.
To provide information on metallicity distribution models of the progenitor satellite, we investigate the metallicity gradient of the progenitor satellite galaxy of the giant stellar stream taking account for negative and positive gradient models. 
%%%%%%%%%%%%%%%%%%%%%%%%%%%%%%%%%%%%%%%%%%%%%%%%%%%%%%%%%%%%%%%%%%%%%%%%%%%%%

%%%%%%%%%%%%%%%%%%%%%%%%%%%%%%%%%%%%%%%%%%%%%%%%%%%%%%%%%%%%%%%%%%%%%%%%%%%%%
A high-resolution $N$-body model is necessary to predict the current metallicity distribution in M31 halo from a metallicity distribution model of the progenitor satellite before the collision. 
To do this, we have performed a high-resolution run of $N$-body simulation for Model A ($M = 3\times 10^9 M_\odot$, $r_{\rm t} = 4.5$ kpc, $c = 0.7$) with $N =$ 524,288. 
Results of the simulation well converge with those of the corresponding low-resolution runs. 
Figure \ref{fig:stream:convergence:metallicity:map} shows spatial metallicity distribution maps with varying metallicity distribution model of the progenitor satellite. 
The top panels in the figure are higher mean metallicity model for the progenitor satellite which assuming mean iron abundance ${\rm [Fe/H]}_{\rm mean}$ of $-0.5$, while the bottom panels exhibit lower mean metallicity model (${\rm [Fe/H]}_{\rm mean} = -1.2$). 
The mean of the observed mass-metallicity relation \citep{dek03} infers that the high and the low metallicity models have the stellar mass of $5\times10^9\, M_\odot$ and $10^8\, M_\odot$, respectively. 
The left and right panels show negative and positive metallicity gradient models ($\Delta {\rm [Fe/H]}$ of $-0.7$ or $0.3$), respectively. 
Figure \ref{fig:stream:convergence:metallicity:map} shows that differences in the metallicity distribution models result in clear differences in the metallicity distribution observed at the present epoch. 
Since particles initially located in the central region of the satellite most likely to exist in the east shell, ${\rm [Fe/H]}$ observed in the east shell region is relatively higher/lower than that observed in other fields for the negative/positive metallicity gradient model. 
Similarly, particles initially located on the outskirt of the progenitor satellite tend to locate on the ``envelope'' of the stream; hence, ${\rm [Fe/H]}$ in the ``envelope'' of the stream becomes lower/higher than that in the ``core'' of the stream for the negative/positive metallicity gradient model. 
%%%%%%%%%%%%%%%%%%%%%%%%%%%%%%%%%%%%%%%%%%%%%%%%%%%%%%%%%%%%%%%%%%%%%%%%%%%%%

%%%%%%%%%%%%%%%%%%%%%%%%%%%%%%%%%%%%%%%%%%%%%%%%%%%%%%%%%%%%%%%%%%%%%%%%%%%%%
In Fig.~\ref{fig:stream:convergence:metallicity:profile}, we compare the above metallicity distribution models and observed metallicity distribution. 
The high metallicity models match with metallicity observed by \citet{guh06, Kalirai2006inner, Kalirai2006outer, gil09} while the low metallicity models are consistent with the metallicity observed by \citet{koc08}. 
The figure shows that the difference of metallicity gradient $\Delta {\rm [Fe/H]}$ of unity, which is corresponding to the observed variety of metallicity gradient \citep{Koleva2009summary}, does not result in a clear difference of metallicity within inner region ($R_{\rm proj} \lesssim 40$ kpc). 
Observations targeting the outer region might distinguish the sign of the metallicity gradient for the progenitor satellite of the giant stream. 
%%%%%%%%%%%%%%%%%%%%%%%%%%%%%%%%%%%%%%%%%%%%%%%%%%%%%%%%%%%%%%%%%%%%%%%%%%%%%

%%%%%%%%%%%%%%%%%%%%%%%%%%%%%%%%%%%%%%%%%%%%%%%%%%%%%%%%%%%%%%%%%%%%%%%%%%%%%
\citet{gil09} reported that the metallicity observed in the ``core'' of the giant stream is $0.17$ dex higher than in the ``envelope''. 
This trend is the same with negative metallicity gradient models; however, the simulations show a difference of only $\sim 0.1$ dex (Fig.~\ref{fig:stream:convergence:metallicity:map}). 
The observed difference of about $0.2$ dex implies a much stronger metallicity gradient compared to observed values for nearby dwarf ellipticals \citep{Spolaor2009, Koleva2009observation, Koleva2009summary}. 
\citet{far12} presented results of spectroscopic measurements of the west shell along the minor axis of the M31 disk. 
They found that the observed metallicity in the west shell was similar to that in the ``core'' of the giant stream. 
This result is consistent with the results presented in this work (Fig.~\ref{fig:stream:convergence:metallicity:map}). 
%%%%%%%%%%%%%%%%%%%%%%%%%%%%%%%%%%%%%%%%%%%%%%%%%%%%%%%%%%%%%%%%%%%%%%%%%%%%%%

%%%%%%%%%%%%%%%%%%%%%%%%%%%%%%%%%%%%%%%%%%%%%%%%%%%%%%%%%%%%%%%%%%%%%%%%%%%%%%
Finally, there are two unexplored and effective ways to constrain the mean iron abundance and the metallicity gradient of the progenitor satellite. 
The first strategy is a direct comparison of metallicity distribution maps given by observations and the simulation (Fig.~\ref{fig:stream:convergence:metallicity:map}). 
Recent photometric observations that cover the stream and the two shells \citep{iba07,mar13} provide information on metallicity. 
The second strategy is focusing on the east shell region. 
As clearly shown in Fig.~\ref{fig:stream:convergence:metallicity:map}, the iron abundance in the east shell region is the highest/lowest for the negative/positive metallicity gradient model. 
Therefore, comparing the metallicity in the east shell and the giant stream would be a possible approach to recovering fossil information on the metallicity distribution of the progenitor satellite. 
Further analysis of photometric observations that cover a wide field and/or future spectroscopic observations focused on the east shell would provide useful clues to investigate the metallicity distribution model of the progenitor. 
%%%%%%%%%%%%%%%%%%%%%%%%%%%%%%%%%%%%%%%%%%%%%%%%%%%%%%%%%%%%%%%%%%%%%%%%%%%%%%
%%%%%%%%%%%%%%%%%%%%%%%%%%%%%%%%%%%%%%%%%%%%%%%%%%%%%%%%%%%%%%%%%%%%%%%%%%%%%%

%%%%%%%%%%%%%%%%%%%%%%%%%%%%%%%%%%%%%%%%%%%%%%%%%%%%%%%%%%%%%%%%%%%%%%%%%%%%%%
%%%%%%%%%%%%%%%%%%%%%%%%%%%%%%%%%%%%%%%%%%%%%%%%%%%%%%%%%%%%%%%%%%%%%%%%%%%%%%
\acknowledgments

We thank M. J. Irwin for allowing us the use of their observational data. 
We are also grateful to an anonymous referee for providing useful information and comments. 
The computations were performed on the FIRST cluster at the Center for Computational Sciences, University of Tsukuba. 
This work was supported in part by JSPS Grants-in-Aid for Scientific Research: (A) (21244013), (C) (25400222) and (S) (20224002), and Grants-in-Aid for Specially Promoted Research by MEXT (16002003).
%%%%%%%%%%%%%%%%%%%%%%%%%%%%%%%%%%%%%%%%%%%%%%%%%%%%%%%%%%%%%%%%%%%%%%%%%%%%%%
%%%%%%%%%%%%%%%%%%%%%%%%%%%%%%%%%%%%%%%%%%%%%%%%%%%%%%%%%%%%%%%%%%%%%%%%%%%%%%

%%%%%%%%%%%%%%%%%%%%%%%%%%%%%%%%%%%%%%%%%%%%%%%%%%%%%%%%%%%%%%%%%%%%%%%%%%%%%%
%%%%%%%%%%%%%%%%%%%%%%%%%%%%%%%%%%%%%%%%%%%%%%%%%%%%%%%%%%%%%%%%%%%%%%%%%%%%%%
\appendix
\section{Fitting formula of $c$ vs $W_{\rm 0}$ in King model}
%%%%%%%%%%%%%%%%%%%%%%%%%%%%%%%%%%%%%%%%%%%%%%%%%%%%%%%%%%%%%%%%%%%%%%%%%%%%%%
We estimate the fitting formulae of the concentration parameter $c$ and the non-dimensional King parameter at the center $W_{\rm 0}$ for our calculation and analysis.
%%%%%%%%%%%%%%%%%%%%%%%%%%%%%%%%%%%%%%%%%%%%%%%%%%%%%%%%%%%%%%%%%%%%%%%%%%%%%%

%%%%%%%%%%%%%%%%%%%%%%%%%%%%%%%%%%%%%%%%%%%%%%%%%%%%%%%%%%%%%%%%%%%%%%%%%%%%%%
The fitting formula for estimating $c$ using given $W_{\rm 0}$ is 
%%%%%%%%%%%%%%%%%%%%%%%%%%%%%%%%%%%%%%%%%%%%%%%%%%%%%%%%%%%%%%%%%%%%%%%%%%%%%%
\begin{equation}
  c=\sum a_i {W_{\rm 0}}^i,
\end{equation}
%%%%%%%%%%%%%%%%%%%%%%%%%%%%%%%%%%%%%%%%%%%%%%%%%%%%%%%%%%%%%%%%%%%%%%%%%%%%%%
where $a_i$ is the expansion coefficient in Table \ref{tab.appendix}. 
The maximum difference between $c$ and our fitting value is 2.6\% ($W_{\rm 0}=0.44$, $c=0.11$) in the region where $0.4 \leq W_{\rm 0} \leq 12$, which corresponds to $0.1 \leq c \leq 2.7$. In the realistic parameter range, the maximum difference is 0.2\% ($W_{\rm 0}=2.5$, $c=0.6$). 
%%%%%%%%%%%%%%%%%%%%%%%%%%%%%%%%%%%%%%%%%%%%%%%%%%%%%%%%%%%%%%%%%%%%%%%%%%%%%%

%%%%%%%%%%%%%%%%%%%%%%%%%%%%%%%%%%%%%%%%%%%%%%%%%%%%%%%%%%%%%%%%%%%%%%%%%%%%%%
The fitting formula for estimating $W_{\rm 0}$ using given $c$ is 
\begin{equation}
W_{\rm 0}=\sum b_i c^i,
\end{equation}
where $b_i$ is the expansion coefficient in Table \ref{tab.appendix}. 
The maximum difference between $W_{\rm 0}$ and our fitting value is 0.6\% ($W_{\rm 0}=0.5$, $c=0.14$) in the region where $0.1 \leq c \leq 2.7$, which corresponds to $0.4 \leq W_{\rm 0} \leq 12$. In the realistic parameter range, the maximum difference is 0.2\% ($W_{\rm 0}=1.9$, $c=0.5$). 
%%%%%%%%%%%%%%%%%%%%%%%%%%%%%%%%%%%%%%%%%%%%%%%%%%%%%%%%%%%%%%%%%%%%%%%%%%%%%%
\begin{deluxetable}{cccccccc}
  \tabletypesize{\scriptsize}
  \tablecaption{Fitting parameters for concentration\label{tab.appendix}}
  \tablewidth{0pt}
  \tablehead{
    \colhead{} & \colhead{$i=0$} & \colhead{$i=1$} & \colhead{$i=2$} & \colhead{$i=3$} & \colhead{$i=4$} & \colhead{$i=5$} & \colhead{$i=6$}
  }
  \startdata
  $a_i$ & $-0.245959$ & $1.20725$ & $-1.24051$ & $0.919125$ & $-0.44859$ & $0.147641$ & $-3.32701\times 10^{-2}$\\
  $b_i$ & $0.274675$ & $0.838297$ & $8.92602$ & $-41.6198$ & $1.6834\times 10^2$ & $-3.53702\times 10^2$ & $4.35371\times 10^2$\\
  \enddata
\end{deluxetable}
%%%%%%%%%%%%%%%%%%%%%%%%%%%%%%%%%%%%%%%%%%%%%%%%%%%%%%%%%%%%%%%%%%%%%%%%%%%%%%
\begin{deluxetable}{ccccccc}
  \tabletypesize{\scriptsize}
  \tablecaption{Fitting parameters for concentration}
  \tablewidth{0pt}
  \tablehead{
    \colhead{} & \colhead{$i=7$} & \colhead{$i=8$} & \colhead{$i=9$} & \colhead{$i=10$} & \colhead{$i=11$} & \colhead{$i=12$}
  }
  \startdata
  $a_{i}$ & $5.1676\times 10^{-3}$ & $-5.50025\times 10^{-4}$ & $3.92371\times 10^{-5}$ & $-1.78697\times 10^{-6}$ & $4.68183\times 10^{-8}$ & $-5.35367\times 10^{-10}$\\
  $b_{i}$ & $-3.42262\times 10^2$ & $1.78236\times 10^2$ & $-61.5313$ & $13.5869$ & $-1.74195$ & $9.88101\times 10^{-2}$\\
  \enddata
\end{deluxetable}
%%%%%%%%%%%%%%%%%%%%%%%%%%%%%%%%%%%%%%%%%%%%%%%%%%%%%%%%%%%%%%%%%%%%%%%%%%%%%%
%%%%%%%%%%%%%%%%%%%%%%%%%%%%%%%%%%%%%%%%%%%%%%%%%%%%%%%%%%%%%%%%%%%%%%%%%%%%%%

%%%%%%%%%%%%%%%%%%%%%%%%%%%%%%%%%%%%%%%%%%%%%%%%%%%%%%%%%%%%%%%%%%%%%%%%%%%%%%
%%%%%%%%%%%%%%%%%%%%%%%%%%%%%%%%%%%%%%%%%%%%%%%%%%%%%%%%%%%%%%%%%%%%%%%%%%%%%%

%%%%%%%%%%%%%%%%%%%%%%%%%%%%%%%%%%%%%%%%%%%%%%%%%%%%%%%%%%%%%%%%%%%%%%%%%%%%%%
%%%%%%%%%%%%%%%%%%%%%%%%%%%%%%%%%%%%%%%%%%%%%%%%%%%%%%%%%%%%%%%%%%%%%%%%%%%%%%

\end{document}